\documentclass[sn-mathphys-num]{sn-jnl}

\usepackage{graphicx}%
\usepackage{multirow}%
\usepackage{amsmath,amssymb,amsfonts}%
\usepackage{amsthm}%
\usepackage{mathrsfs}%
\usepackage[title]{appendix}%
\usepackage{xcolor}%
\usepackage{textcomp}%
\usepackage{manyfoot}%
\usepackage{booktabs}%
\usepackage{algorithm}%
\usepackage{algorithmicx}%
\usepackage{algpseudocode}%
\usepackage{listings}%

\usepackage{lmodern}
\usepackage{anyfontsize}
\usepackage{subfigure} 
\usepackage[version=4]{mhchem} 
\usepackage{braket}
\usepackage{amssymb}  
\usepackage{svg}
\usepackage{bm}
\usepackage{float}
\usepackage[T1]{fontenc}
\usepackage{amsmath}
\usepackage[all]{nowidow}

\begin{document}

\title[Article Title]{Tutorial: From Topology to Hall Effects -- Implications of Berry Phase Physics}


\author*[1]{\fnm{Nico} \sur{Sprinkart}}\email{nico.sprinkart@uni-konstanz.de}

\author*[1]{\fnm{Elke} \sur{Scheer}}\email{elke.scheer@uni-konstanz.de}

\author*[1]{\fnm{Angelo} \sur{Di Bernardo}}\email{angelo.dibernardo@uni-konstanz.de}

\affil[1]{\orgdiv{Department of Physics}, \orgname{University of Konstanz}, \orgaddress{\street{Universitätstraße 10}, \city{Konstanz}, \postcode{78464}, \country{Germany}}}




\abstract{The Berry phase is a fundamental concept in quantum mechanics with profound implications for understanding topological properties of quantum systems. This tutorial provides a comprehensive introduction to the Berry phase, beginning with the essential mathematical framework required to grasp its significance. We explore the intrinsic link between the emergence of a non-trivial Berry phase and the presence of topological characteristics in quantum systems, showing the connection between the Berry phase and the band structure as well as the phase's gauge-invariant nature during cyclic evolutions. The tutorial delves into various topological effects arising from the Berry phase, such as the quantum, anomalous, and spin Hall effects, which exemplify how these quantum phases manifest in observable phenomena. We then extend our discussion to cover the transport properties of topological insulators, elucidating their unique behaviour rooted in Berry phase physics. This tutorial aims at equipping its readers with a robust understanding of the basic theory around the Berry phase and its pivotal role in the realm of topological quantum phenomena.}




\maketitle
\newpage
\section{Introduction}\label{section_Topology}

Quantum mechanics is one of the fundamental theories at the foundation of modern physics. This theory describes physical systems at the smallest length and time scales, where the laws of classic theories like Newtonian mechanics do not hold anymore. Quantum mechanics properly considers that the positions and velocities of atomic and subatomic particles are not exactly defined due to wave-particle duality, which in turn implies that the state of a quantum mechanical system can only be expressed in terms of probabilities. To fulfil this description, complex wave functions $\psi$ are introduced. The squared modulus of $\psi$ represents the probability density of a quantum system to be in a specific state $n$. The fundamental equation describing the time evolution of a quantum mechanical system is the time-dependent Schrödinger equation 
\begin{align}
    \hat{H}\psi = i\hbar \frac{\partial}{\partial t}\psi,
    \label{eq_Schrödinger}
\end{align}
where $\hat{H}$ is the Hamiltonian, $i$ the imaginary unit, $\hbar$ the reduced Planck's constant and $t$ the time. Often, the wave function $\psi$, which is solving the corresponding Schrödinger equation \eqref{eq_Schrödinger} with the appropriate boundary conditions, has a sinusoidal dependence of its phase. Although a phase shift of $\psi$ does not affect the probability density $|\psi|^2$ of finding the system in a specific state, the phase of $\psi$ carries information on the interference and superposition properties of different quantum states. The phase consists of a "dynamic" part, which depends on the temporal evolution of the energy $E$ of the state, and of a time-independent "geometric" part. 

\indent In 1928, during the early stages of quantum mechanics, Fock showed that the geometric part of the phase can be set to zero by a gauge transformation for non-cyclic evolutions \cite{Fock1928}. This practice was widely accepted and, for the following 50 years, the geometric phase was mostly assumed to be unimportant \cite{Bohm2013}, until in 1983 Berry reconsidered the influence of cyclic evolutions, namely evolutions where a quantum state is periodically evolving in time \cite{Berry1984}. Berry found that, in this configuration, the geometric phase can yield a more general explanation of the Aharonov-Bohm effect \cite{Aharonov1959} than previous models. Soon after, the scientific community realised that the geometric phase can also explain other effects such as the anomalous velocity \cite{Luttinger1954} or the Pancharatnam phase \cite{Pancharatnam1956}. Towards the end of the 1980s, a variety of experiments were performed which demonstrated that the geometric phase can affect the properties of different quantum systems \cite{Suter1988, Tycko1987, Tomita1986}. Today, it is well-established that the geometric phase, now also known as Berry phase, plays a key role in the field of topological physics and contributes to the theories of topological materials \cite{Hasan2010, Qi2011}, adiabatic quantum computing \cite{Albash2018}, spintronics \cite{Fabian2004}, optics \cite{BHANDARI1997}, photonics \cite{Jisha2021}, among others.

\indent In this tutorial, we first introduce the basic mathematical formalism needed to understand the Berry phase. We show that a Fock gauge transformation \cite{Fock1928} does not affect the Berry phase for cyclic evolutions. The tutorial then illustrates how the occurrence of a non-trivial Berry phase is intrinsically connected with the presence of topological properties in a quantum system. A variety of topological effects stemming from the Berry phase, like the quantum and the anomalous Hall effect, are then introduced, followed by a description of the main transport properties of topological insulators.

\section{Fundamentals of Berry Phase Physics}\label{Section_FundamentalsBerryPhase}
Let us consider a system with a time-dependent Hamiltonian $\hat{H}(t)$ acting on its wave function $\ket{\psi(t)}$. At $t=0$ the system may be in an initial state $\ket{\psi(0)}=\ket{n(0)}$, which is a so-called "instantaneous eigenstate" of $\hat{H}(0)$ with a corresponding eigenvalue $E_n(0)$. The instantaneous eigenstates of the system may be normalised such that
\begin{align}
\label{eq.InstEigenstatesNormalized}
    \forall t:\quad\braket{n(t)|m(t)}=\delta_{nm},
\end{align}
where $\delta_{nm}$ is the Kronecker symbol. If the system evolves in time slowly enough and the Hamiltonian spectrum fulfils the condition
\begin{align}
\label{eq.looololol}
\forall t\in[0,T]: \quad ...\leq E_{n-1}(t)<E_n(t)<E_{n+1}(t)\leq...\,,
\end{align}
meaning that at any given time, in a certain time period of duration $T$, the $n$-th eigenvalue is separated from the rest of the spectrum, then the adiabatic theorem \cite{Born1928, Kato1950} states that the $n$-th instantaneous eigenstate evolves with continuity, remaining in the $n$-th eigenstate at any time instant \cite{Suzuki2018}. In other terms, the system does not jump into another state, but each eigenstate and the corresponding eigenvalue change continuously over time. 

\begin{figure}[H]
  \centering
  \includegraphics[width=0.9\textwidth]{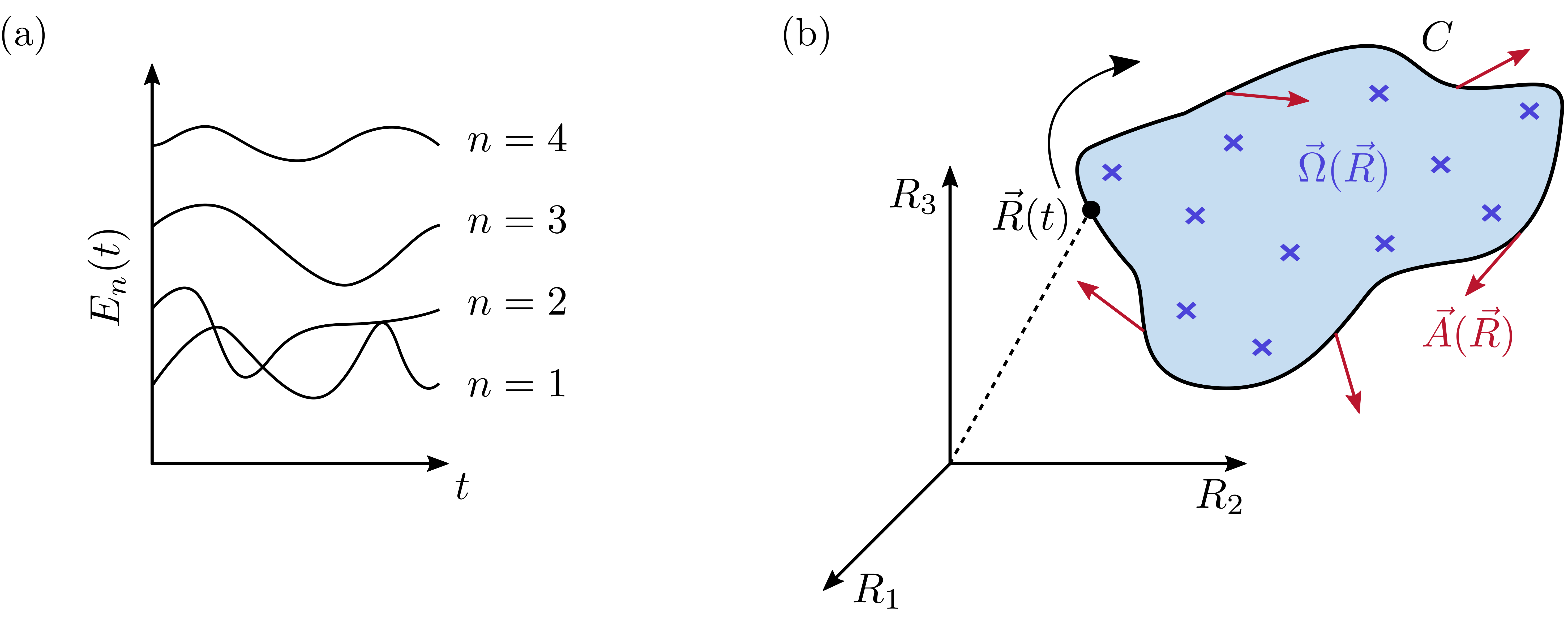}
  \caption[Berry connection and curvature]{Adiabatic, cyclic evolution of quantum systems. (a) Energy spectrum of a Hamiltonian $\hat{H}(\vec{R}(t))$ over time. The states $n=3,4$ satisfy Equation \eqref{eq.looololol}, while the states $n=1,2$ show degeneracy and are therefore not suitable for the adiabatic approximation. (b) A parameter $\vec{R}$ being traced along a closed loop $C$ in parameter space. The resulting Berry connection $\vec{A}(\vec{R})$ constitutes a vector field that is defined along $C$ (red arrows). $\vec{A}(\vec{R})$ generates a flux through the surface $S$ enclosed by $C$ (light blue), and the corresponding flux density $\vec{\Omega}(\vec{R})$ is the Berry curvature (blue crosses).} 
  \label{fig_BerryCurvature}
\end{figure}

\indent In Figure \ref{fig_BerryCurvature}(a) this condition is fulfilled for $n=3,\,4$ and broken for $n=1,\,2$. Such a time-dependent adiabatic transport is given, for example, when a voltage bias is applied to a quantum system like in the Hall regimes described in Chapter \ref{Section_AHEundSHE}. The corresponding instantaneous eigenvalue equation is
\begin{align}
\label{eq.InstantEigenvalueEquation}
    \hat{H}(t)\ket{n(t)}=E_n(t)\ket{n(t)},
\end{align}
which means that we can find at any time a state $\ket{n(t)}$ that fulfils the stationary Schrödinger equation. To study how the system described by $\hat{H}(t)$ evolves, one needs to determine the solutions of the time-dependent Schrödinger equation \eqref{eq_Schrödinger}. The instantaneous eigenstates are not sufficient to solve Equation \eqref{eq_Schrödinger}, but they can be used as auxiliary states in the ansatz
\begin{align}
\label{eq.Ansatz}
    \ket{\psi}=e^{-i\phi(t)}\ket{n(t)},
\end{align}
where $\phi(t)$ is the time-dependent phase. Usually, the operator $\hat{H}$ does not contain a time derivative. Then, after substituting Equation \eqref{eq.Ansatz} into the time-dependent Schrödinger equation \eqref{eq_Schrödinger}, calculating the time derivative of $\ket{\psi}$, and simplifying the resulting equation, one finds  
\begin{align}
\label{eq.Schritt1}
    \frac{\partial \phi(t)}{\partial t}\ket{n(t)}+i\frac{\text{d}}{\text{d}t}\ket{n(t)}=\frac{1}{\hbar}\hat{H}(t)\ket{n(t)}.
\end{align}
Multiplying Equation \eqref{eq.Schritt1} from the left by $\bra{n(t)}$,\footnote{This is only valid in the adiabatic approximation, otherwise the inner product should be calculated with any state in the Hilbert space and not only with $\bra{n(t)}$.} applying Equations  \eqref{eq.InstEigenstatesNormalized} and \eqref{eq.InstantEigenvalueEquation} and solving for $\dot{\phi}(t)$ results in
\begin{align}
\label{eq.Schritt2}
    \dot{\phi}(t)=\frac{1}{\hbar}E_n(t)-i\bra{n(t)}\frac{\text{d}}{\text{d}t}\ket{n(t)}.
\end{align}
We use dotted parameters, e.g., $\dot{\phi}(t)$, to represent the time derivative. Finally, through integration over time of Equation \eqref{eq.Schritt2}, it is possible to derive a more general expression for the phase $\phi(t)$:
\begin{align}
\label{eq.dyn+BerryPhase}
    \phi(t)=\underbrace{\frac{1}{\hbar}\int_{0}^t E_n(t')\,\text{d}t'}_{\substack{\text{dynamical phase}}}
    -\,\,\,\underbrace{i\int_{0}^t \bra{n(t')}\frac{\text{d}}{\text{d}t'}\ket{n(t')} \,\text{d}t'}_{\substack{\text{Berry phase }\gamma(t)}}.
\end{align}
The first term of Equation \eqref{eq.dyn+BerryPhase} is the dynamical phase of the wave function and depends on the state's energy and on the time. The second term is the geometric phase $\gamma(t)$, to which we refer in the following as Berry phase, under the implicit assumption that the adiabatic approximation holds. A generalised yet intricate formulation of the geometric phase is the Aharonov-Anandan phase that can also be used beyond the adiabatic approximation \cite{Anandan1987}. 
\par So far, it has been assumed that the Hamiltonian $\hat{H}(t)$, describing the system, has an explicit time dependence. The time dependence, however, usually originates from an external parameter $\vec{R}(t)$ such that $\hat{H}=\hat{H}(\vec{R}(t))$ and $\ket{n}=\ket{n(\vec{R}(t))}$. As already mentioned, and as will be demonstrated later, the geometric phase can be set to zero by a gauge transformation for non-cyclic evolutions \cite{Fock1928}. Therefore, in the following we only consider cyclic evolutions, which implies that $\vec{R}$ evolves along a closed path $C$ in parameter space within the time interval $[0,T]$, so that $\vec{R}(T)=\vec{R}(0)$. This condition also means that the Hamiltonian returns to its initial form after one loop, meaning that $\hat{H}(T)=\hat{H}(0)$, but the same does not necessarily apply to the wave function $\ket{\psi}$ \eqref{eq.Ansatz}. The absolute value of $\ket{\psi(T)}$ coincides with the initial absolute value of $\ket{\psi(0)}$, but depending on the exact geometry of the travelled path $C$, the Berry phase can change during the transport.\par

\begin{figure}[H]
  \centering
  \includegraphics[width=0.9\textwidth]{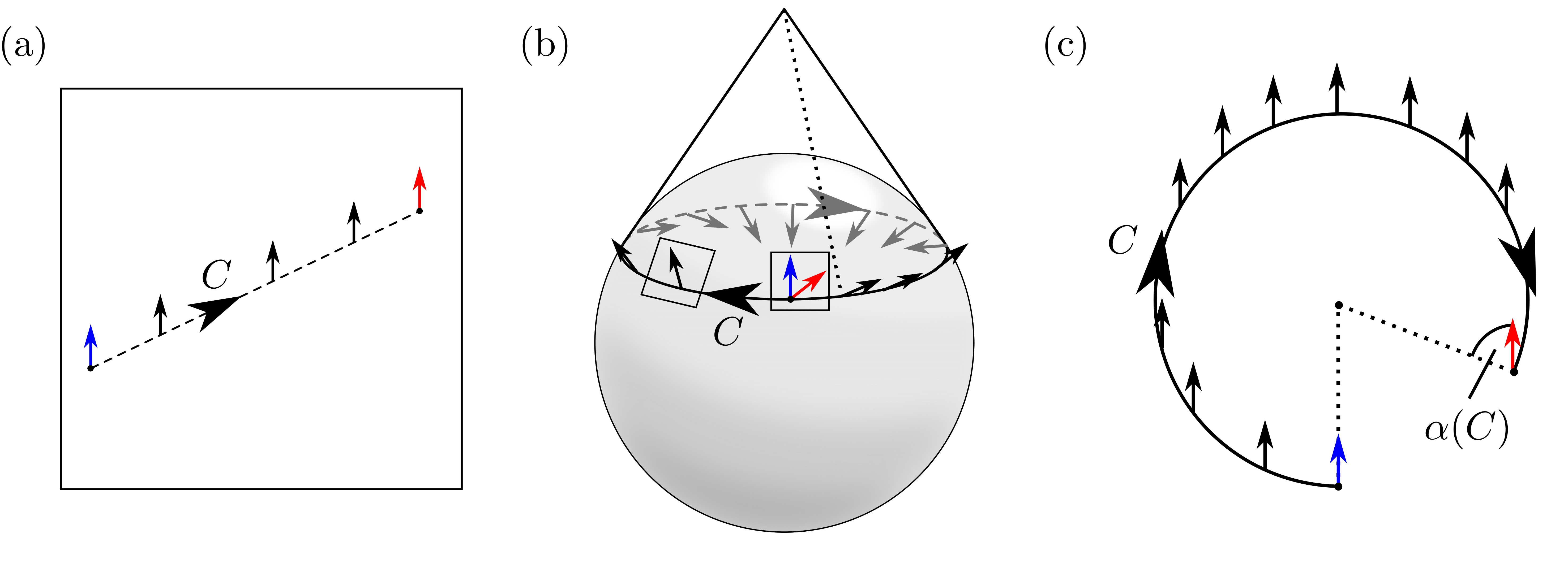}
  \caption[Parallel transport in Euclidean and curved space]{Parallel transport of a vector $\vec{r}$ along a path $C$. The vectors before, during and after transport are printed in blue, black, and red, respectively. (a) Parallel transport in Euclidean space. Vectors associated to all intermediate states are parallel to each other. (b) Parallel transport along a fixed latitude on the surface of a sphere. The vector rotates inside the planes tangential to the path $C$. The set of all tangential planes forms a cone on top of the sphere. (c) Flattened cone of tangential planes. In this representation, all intermediate states are parallel to each other, but the final vector is rotated by an angle $\alpha(C)$, relative to the latitude.}
  \label{fig_ParallelTransport}
\end{figure}

To better understand how the phase can change depending on the specific path considered, we draw an analogy with the transport of a classic vector $\vec{r}$ in real space. Figure \ref{fig_ParallelTransport}(a) shows the transport of $\vec{r}$ along a path $C$ in Euclidean space. In this case, the tip and tail of the vector are each shifted by the same distance and along the same direction during the transport process. Consequently, the initial vector (in blue), all its intermediate states during transport (in black), and the final vector (in red) are parallel to each other. This fact, however, is not met anymore when $\vec{r}$ is transported along a fixed latitude on the surface of a sphere, as shown in Figure \ref{fig_ParallelTransport}(b).
The surface of the sphere is a curved space, also called manifold. The area near each point of a manifold can be locally resembled by a 
Euclidean subspace. In the case of the sphere, a set of possible Euclidean subspaces are the tangential planes along the path $C$. Two representative planes are illustrated in Figure \ref{fig_ParallelTransport}(b). When the vector transport occurs along a fixed latitude, the sum of all tangential planes constitutes a cone sitting on top of the sphere [Figure \ref{fig_ParallelTransport}(b)]. For each tangential plane associated with an infinitesimal shift of $\vec{r}$ along $C$, the tip and tail of $\vec{r}$ are each shifted by the same distance and direction. However, when the vector is observed from the outside, tip and tail are moving along different latitudes (and longitudes), which causes a rotation of the vector. If we flatten the cone of tangential planes [Figure \ref{fig_ParallelTransport}(c)], we see that, in this representation, $\vec{r}$ remains parallel to its initial direction at all times. Such a transport process is called "parallel transport". But in both representations we see that the vector $\vec{r}$ has rotated by an angle $\alpha(C)$, which depends on the exact geometry of the path $C$, after returning to its initial position. A real-world example of such a transport process is given by the Foucault pendulum, where the plane of oscillation rotates with respect to the external environment, while it is transported along a fixed latitude by the rotation of the Earth.
\par The wave function $\ket{\psi(\vec{R})}$ can experience something similar, noticeable as a change of its Berry phase, when it is transported through the parameter space. In analogy with the example made above, the change of the Berry phase depends on the geometry of the loop but not on the time. In other terms, the change in Berry phase occurs independently of whether the system completes the loop slower or faster. We can get to the same conclusion also analytically by calculating the time derivative of the Berry phase from Equation \eqref{eq.dyn+BerryPhase}, which yields: 
\begin{align}
\label{eq.TimeDerivative}
    \frac{\text{d}}{\text{d}t}\ket{n(\vec{R}(t))}=\frac{\text{d}\ket{n(\vec{R}(t)}}{\text{d}\vec{R}}\cdot \frac{\text{d}\vec{R}}{\text{d}t}=\nabla_{\vec{R}}\cdot\ket{n(\vec{R}(t)} \cdot\frac{\text{d}\vec{R}}{\text{d}t}.
\end{align}
By inserting the result of Equation \eqref{eq.TimeDerivative} into Equation \eqref{eq.dyn+BerryPhase}, we can drop the time dependence and define $\gamma(t)$ as a curve integral over the closed path $C$ in parameter space:
\begin{align}
\label{eq.BerryVectorPotential}
    \gamma(t)=\int_{0}^t i\bra{n(\vec{R}(t'))}\nabla_{\vec{R}}\ket{n(\vec{R}(t'))} \frac{\text{d}\vec{R}}{\text{d}t'}\cdot\text{d}t'= \oint_C \underbrace{i\bra{n(\vec{R})} \nabla_{\vec{R}} \ket{n(\vec{R})}}_{\substack{:=\,\vec{A}(\vec{R})}} \cdot\,\text{d}\vec{R}=\gamma(C).
\end{align}
The integrand $\vec{A}(\vec{R})$ of Equation \eqref{eq.BerryVectorPotential} is known as the "{Berry} connection" and can be seen as a kind of vector field defined along the path $C$ [see Figure \ref{fig_BerryCurvature}(b)]. Equation \eqref{eq.BerryVectorPotential} can be rewritten by applying Stokes' theorem, which states that the integral of the vector field $\vec{A}$ along the loop $C$ equals the flux generated by $\vec{A}$ through the surface $S$ enclosed by $C$:
\begin{align}
\label{eq.BerryCurvature}
\gamma(C)=\oint_C \vec{A}(\vec{R})\cdot \text{d}\vec{R}\,\, 
\overset{\textrm{Stokes}}{=} \int_S \underbrace{\nabla_{\vec{R}}\times\vec{A}(\vec{R})}_{\substack{:=\,\vec{\Omega}(\vec{R})}} \cdot\,\text{d}\vec{S}.
\end{align}
The integrand $\vec{\Omega}(\vec{R})$ of Equation \eqref{eq.BerryCurvature} is the Berry curvature. The Berry curvature is the curl of the vector field $\vec{A}(\vec{R})$ and can be seen as a magnetic flux density defined in momentum space [see Figure \ref{fig_BerryCurvature}(b)]. This analogy to electrodynamics helps to understand the role of the Berry curvature in topological problems and will be useful in the derivations reported below.
\par We now discuss how the Berry phase behaves under a gauge transformation. The Berry connection $\vec{A}(\vec{R})$ can be arbitrarily gauge-transformed by adding the divergence of a vector field $\chi(\vec{R})$ which can be freely chosen, similar to the gauge transformation for the magnetic vector potential used in electrodynamics:
\begin{align}
\label{eq.BerryPhaseGauge}
    \vec{A}'(\vec{R})=\vec{A}(\vec{R})+\nabla_{\vec{R}}\cdot\chi(\vec{R}).
\end{align}
The corresponding Berry phase $\gamma'$ for an arbitrary path between an initial position $\vec{R}_i$ and a final position $\vec{R}_f$ is then given by
\begin{align}
\label{eq.BerryPhaseOfAnyPath}
    \gamma'=\int_{\vec{R}_i}^{\vec{R}_f} \vec{A}(\vec{R})\cdot \text{d}\vec{R}+\int_{\vec{R}_i}^{\vec{R}_f} \nabla_{\vec{R}}\chi(\vec{R})\cdot\text{d}\vec{R}=
    \gamma+\underbrace{\chi(\vec{R}_f)-\chi(\vec{R}_i)}_{\substack{:=\,\Delta\chi}}.
\end{align}
Based on Equation \eqref{eq.BerryPhaseOfAnyPath}, Fock argued that it is always possible to choose the function $\chi(\vec{R})$, such that $\Delta\chi$ compensates the Berry phase $\gamma$ \cite{Fock1928}. As already explained, this statement is generally valid except for closed paths, for which $\vec{R}_f=\vec{R}_i$. In this case, the condition that wave functions must fulfil single-valuedness, i.e., the value of a wave function at any point in space must be unique and well-defined, gives
\begin{align}
e^{i\chi(\vec{R}_f)}=e^{i\chi(\vec{R}_i)}\text{\quad and \,\,} \chi(\vec{R}_f)= \chi(\vec{R}_i) +2\pi\cdot k\text{,\quad with } k\in\mathbb{N}_0.
\end{align}

\noindent We conclude that the Berry phase is gauge-invariant for closed paths in parameter space, which is the case when a quantum mechanical system evolves periodically in time. This can be the case for e.g. crystalline solids.\par
\begin{figure}[H]
  \centering
  \includegraphics[width=0.9\textwidth]{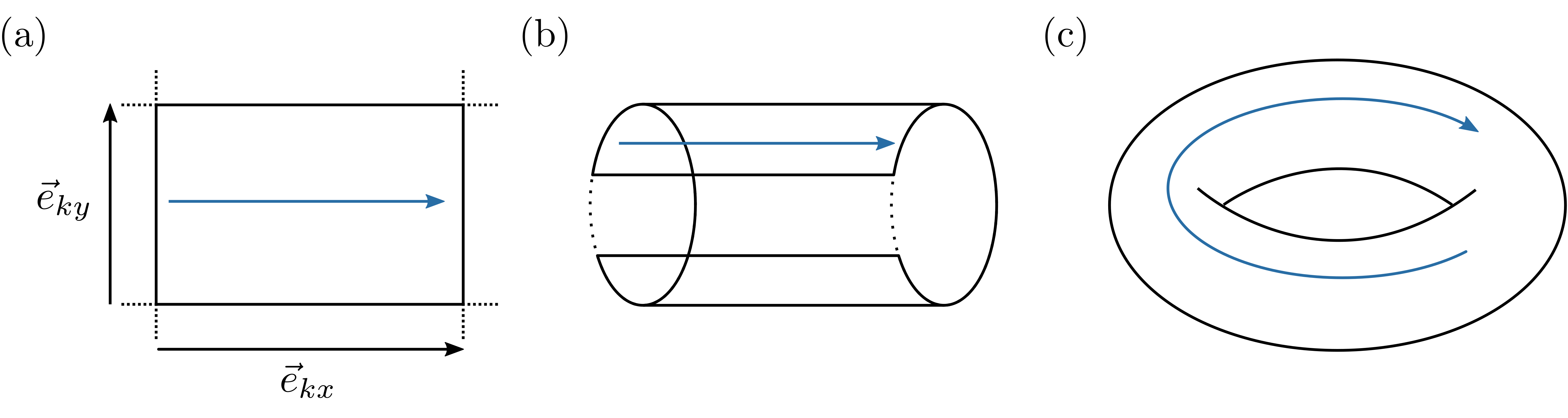}
  \caption[Topology of reciprocal space]{Topology of reciprocal space, where a wave vector $\vec{k}$ is travelling in $x$-direction. (a) When $\vec{k}$ passes from one unit cell to another, it appears as if $\vec{k}$ entered the same cell again, due to the periodicity of the reciprocal lattice and the indistinguishability of its unit cells. Therefore, the infinite reciprocal space can be modelled as a finite space. By connecting the unit cells along their sides parallel to $\vec{e}_{kx}$ first (b) and then along their sides parallel to $\vec{e}_{ky}$ (c), it can be inferred that the as-constructed reciprocal space has the topology of a torus. In this representation, the vector $\vec{k}$ is therefore travelling along a closed path.} 
  \label{fig_Torus}
\end{figure}

In condensed matter physics, the movement of an electron through crystalline solids can be described assuming a periodical potential $V(\vec{r})=V(\vec{r}+\vec{e}_i)$ felt by the electron, where $\vec{e}_i$ is a lattice vector. For such a system, Bloch's theorem \cite{Bloch1929} states that the wave functions of the electrons can be written as the product of a plane wave times a periodic function $u(\vec{r})=u(\vec{r}+\vec{e_i})$ that has the same periodicity as the crystal lattice. For the $n$-th energy band the wave function is then given by 
\begin{align}
\label{eq.BlochsTheorem}
    \psi_n(\vec{k},\vec{r})=e^{i\vec{k}\cdot\vec{r}}u_n(\vec{r}),
\end{align}
where $\vec{r}$ is the position in real space and $\vec{k}$ the wave vector that is defined in reciprocal space. The reciprocal lattice of the system has also a certain periodicity such that the condition $\psi_n(\vec{k})=\psi_n(\vec{k}+\vec{e}_{ki})$ holds, with $\vec{e}_{ki}$ being a reciprocal lattice vector. This condition implies that, if $\vec{k}$ travels in reciprocal space and goes from one unit cell to another, from the point of view of $\vec{k}$, it appears as if $\vec{k}$ entered the same unit cell again, since all unit cells are indistinguishable from each other. As a result, the Brillouin zone (BZ) (i.e., the primitive lattice cell in reciprocal space) has the topology of a torus, with $\vec{k}$ running along a closed path on its surface. This concept is illustrated in Figure \ref{fig_Torus}.
\noindent Figure \ref{fig_Torus} shows that the movement of a wave vector $\vec{k}$ through a periodic reciprocal space is equivalent to that of $\vec{k}$ moving along a closed path on a torus. For such a system, we can define a gauge-invariant Berry phase. 
By solving the Schrödinger equation with $\psi_n(\vec{k})$ satisfying the conditions given by Bloch's theorem (Equation \eqref{eq.BlochsTheorem}) for a periodic potential $V(\vec{r})$, we can determine the energy eigenvalues $E_n(\vec{k})$ corresponding to the eigenstates of the systems. The energy eigenvalues in dependence of the wave vector $\vec{k}$ with $E_n(\vec{k})=E_n(\vec{k}+\vec{e}_{ki})$ constitute the band structure of the system, which is closely related to the Berry curvature $\vec{\Omega}$. This result can be obtained by rewriting Equation \eqref{eq.BerryCurvature}, as demonstrated in the \hyperref[section-addendum]{Appendix}, where all the steps of the derivation are reported. Here, we only report the final result, which is 
\begin{align}
\label{eq.crazyFormel}
     \vec{\Omega}(\vec{R})=i\sum_{m\neq n}\frac{\bra{n(\vec{R})}\nabla_{\vec{R}}\hat{H}(\vec{R})\ket{m(\vec{R})}\times\bra{m(\vec{R})}\nabla_{\vec{R}}\hat{H}(\vec{R})\ket{n(\vec{R})}}{\left(E_n(\vec{R})-E_m(\vec{R})\right)^2}.
\end{align}
At first sight, Equation \eqref{eq.crazyFormel} looks quite intricate, but it becomes easier to understand when considering its parts individually. The numerator of Equation \eqref{eq.crazyFormel} is a measure for the strength and direction of the interaction between two states, $\ket{n}$ and $\ket{m}$, through the spatial variation of the Hamiltonian. In other words, the numerator describes how the eigenstates of the system are influenced by spatial changes in the Hamiltonian. The more crucial part lies in the denominator. It shows that the largest contributions to the Berry curvature, which is defined for every point in the parameter space, arise at near-degeneracies in the band structure. Therefore, the band structure can be interpreted as the origin of the Berry curvature. If the system describes a loop through the Berry curvature landscape, it accumulates a certain Berry phase.
\par In the case of degenerate eigenstates, Equation \eqref{eq.crazyFormel} exhibits a singularity that corresponds to a monopole in the parameter space. At such singularity points, Equation \eqref{eq.crazyFormel} can be replaced by the more intricate non-abelian Berry curvature \cite{Wilczek1984}. The number of monopoles inside the BZ defines the so-called Chern number $Q_n$. For the $n$-th band, the Chern number can be determined by using the residue theorem:
\begin{align}
\label{eq.ChernNumber}
Q_n=\frac{1}{2\pi i}\int_\textsc{BZ}\Omega_n (\vec{k})\quad\text{d}\vec{k}.
\end{align}
$Q_n$ is an intrinsic property of the band structure and it is connected with a variety of topological effects, including the quantisation of the Hall conductance explained in section \ref{section_QHE}. Further details about the topics covered in this chapter can be found in refs. \cite{Bohm2013, Berry1984, Xiao2010, Zwiebach2010, Manoukian2007, Ong2005}.

\section{Classical and Quantum Hall Effect}
\label{section_QHE}
In the previous chapter, we discussed the origin of the Berry curvature $\vec{\Omega}$ and its significance for quantum mechanical systems. The Berry curvature $\vec{\Omega}$ is also responsible, among other aspects, for the quantisation of the Hall conductance in strong magnetic fields, which is the main topic of this chapter.
\par
\begin{figure}[H]
  \centering
 \includegraphics[width=0.9\textwidth]{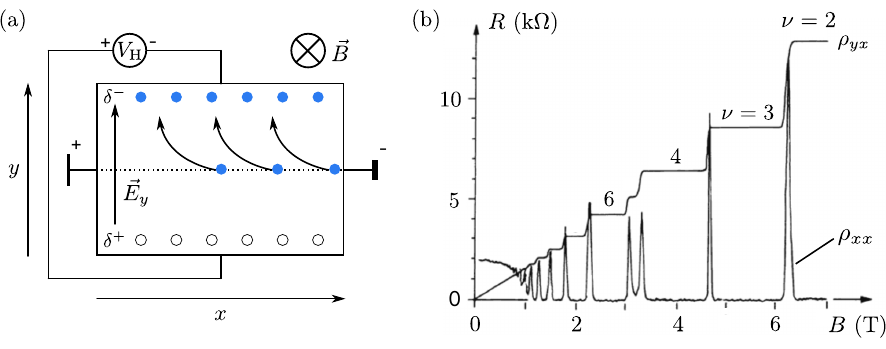}
  \caption[Classical and quantum Hall effects.]{Classical and quantum Hall effects. (a) A current flowing along the $x$ direction, under a magnetic field $B_z$, leads to a separation of charges $\delta$ along the $y$ direction, which in turn induces a transversal Hall voltage $V_\textrm{H}$. (b) Magnetic-field dependence of the longitudinal resistance $\rho_{xx}$ (bottom curve) and of the transversal resistance $\rho_{yx}$ (top curve). Figure adapted from ref. \cite{RoyalAcademy1998}. In agreement with Equation \eqref{eq.QuantizedHallResistivity}, the Hall resistivity $\rho_{yx}$ forms plateaus, when the Fermi energy $E_\textrm{F}$ falls within the gap between two Landau levels [see also Figures \ref{fig_LandauLevel}(d)-(f)]. In this case, charge transport occurs along dissipationless edge channels and the longitudinal resistance $\rho_{xx}$ vanishes. When passing from one plateau to another, the system goes through a metallic phase, which results in spikes in $\rho_{xx}$.} 
  \label{fig_QHE}
\end{figure}

 To introduce the Hall effect, we consider a metallic plate with lateral sizes $L_x$ and $L_y$ and thickness $d$. If an electric field $\vec{E}$ is applied along the $x$-direction, $\vec{E}$ induces an electric force $\vec{F}_E=q\cdot\vec{E}$ onto the charge carriers, which results in the flow of charge current along the $x$-direction with current density $\vec{J}$. For isotropic materials, this relationship can be described by Ohm's law $\vec{J}=\sigma\cdot\vec{E}$, where $\sigma$ is the conductivity of the material. 
\noindent If a magnetic field $\vec{B}$ is then applied along the $z$-direction, the charge current acquires an additional $y$-component due to the deflection of the charge carriers by the Lorentz force $\vec{F}_L=q\cdot\vec{v}\times\vec{B}$. The conductivity therefore becomes directional, and Ohm's law has to be generalised for the multi-dimensional space (two-dimensional for the case under consideration) using the conductivity tensor $\bm{\sigma}$:
\begin{align}
    \begin{bmatrix} J_x \\ J_y\end{bmatrix} =
    \begin{bmatrix} \sigma_{xx} & \sigma_{xy} \\ \sigma_{yx} & \sigma_{yy} \end{bmatrix}
    \begin{bmatrix} E_x \\ E_y \end{bmatrix},
\end{align}
Each component of the conductivity tensor $\sigma_{nm}$ describes how an electric field along the $m$-direction generates a charge current along the $n$-direction. Due to rotational symmetry about the $z$-axis for the considered system, $\sigma_{xx}=\sigma_{yy}$ and $\sigma_{xy}=-\sigma_{yx}$. These correlations become evident when considering Figure \ref{fig_QHE}(a). The sign of the off-diagonal elements of the tensor $\bm{\sigma}$ depends on the direction of $B_z$. The inverse of $\bm{\sigma}$ is the resistivity tensor $\bm{\rho}=\bm{\sigma}^{-1}$. $\bm{\sigma}$ and $\bm{\rho}$ fulfil the condition $\bm{\sigma}\bm{\rho}=I_2$, where $I_2$ is the two-dimensional identity matrix. As a result, the components of $\bm{\sigma}$ can be also expressed as
\begin{align}
\label{eq.NougatBits}
\sigma_{xx}=\sigma_{yy}=\frac{\rho_{xx}}{\rho_{xx}^2+\rho_{xy}^2} \quad\text{and}\quad
\sigma_{xy}=-\sigma_{yx}=\frac{-\rho_{xy}}{\rho_{xx}^2+\rho_{xy}^2}.
\end{align}
Vice versa, the components of $\bm{\rho}$ are given by simply exchanging $\rho$ and $\sigma$ in Equation \eqref{eq.NougatBits}. 
\par Since real systems have a finite size, the $y$-component of the current for a system like that shown in Figure \ref{fig_QHE}(a) always generates opposite charge accumulation at the top and bottom edges of the metallic plate. This charge accumulation increases until the built-up electric field $E_y$ reaches a certain value, at which the electric force and the Lorentz force compensate each other, meaning that:
\begin{align}
\label{eq.Equilibrium}
    q\cdot E_y=q\cdot v_x\cdot B_z.
\end{align}

Once this equilibrium condition is reached, the current again flows strictly along the $x$-direction and $I_y=0$. Using Equation \eqref{eq.Equilibrium} and the expression for the current density $J_x=n_e\cdot q\cdot v_x=I_x/(L_y\cdot d)$, where $n_e$ is the charge carrier density, we can derive an expression for the Hall voltage $V_\textrm{H}$ that is measurable between the top and bottom edges of the system:
\begin{align}
\label{eq.classicHallVoltage}
    V_\textrm{H}=E_y\cdot L_y=v_x\cdot B_z\cdot L_y=\underbrace{\frac{1}{n_e\cdot q}}_{\substack{:=A_\textrm{H}}}J_x\cdot B_z \cdot L_y=A_\textrm{H}\frac{I_xB_z}{d}.
\end{align}
The proportionality factor $A_\textrm{H}$ is the Hall constant, and it is a material-specific parameter.

\par We can also define the Hall resistivity $\rho_{yx}$, which corresponds to the $yx$-component of the resistivity tensor $\bm{\rho}$, as
\begin{align}
\label{eq.ClassicHallResistivity}
    \rho_{yx}=\frac{E_y}{J_x}=\frac{v_x\cdot B_z}{n_e\cdot q\cdot v_x}=A_\textrm{H}\cdot B_z.
\end{align}
We note here that in a two-dimensional (2D) system resistivity and (sheet) resistance are equal, i.e. also have the same dimension and unit. This also holds for for the Hall resistivity and the Hall resistance. Equations \eqref{eq.classicHallVoltage} and \eqref{eq.ClassicHallResistivity} describe the classical Hall effect, in which the transversal Hall voltage $V_\textrm{H}$ is linearly dependent on the current density $J_x$, and both $V_\textrm{H}$ and $\rho_{yx}$ depend linearly on the strength of the applied magnetic field $B_z$. However, the latter condition is only valid for low applied $B_z$, as shown in Figure \ref{fig_QHE}(b) for $B_z< 1$\,T.

\par
\begin{figure}[h]
  \centering
   \includegraphics[width=0.9\textwidth]{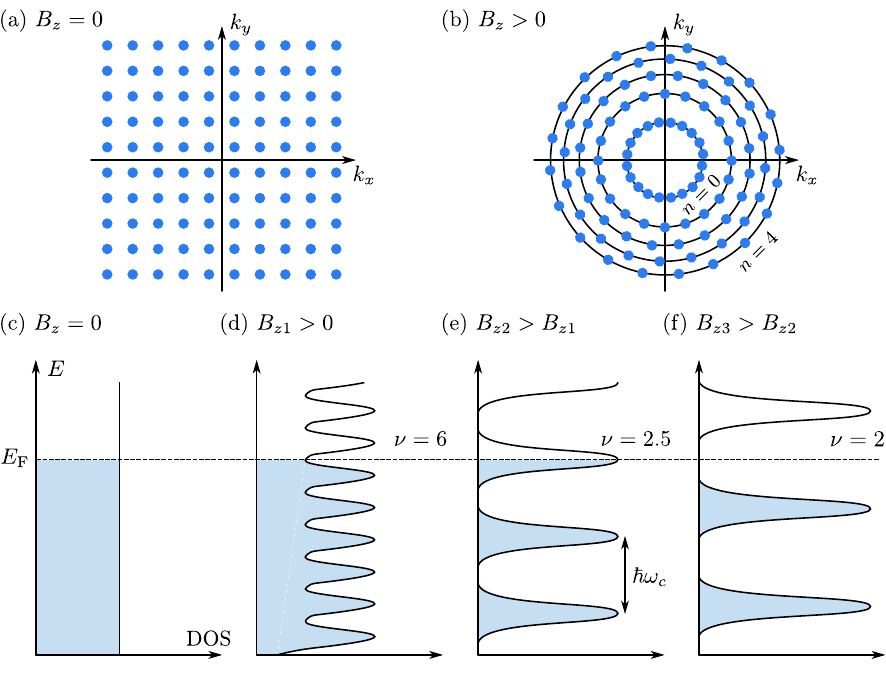}
  \caption[Effects of a magnetic field on a 2DEG]{Effects of an applied magnetic field $B_z$ on the DOS of a 2DEG. (a) Energy states with uniform distribution in $k$-space. (b) Once $B_z$ is applied to the system in (a), $B_z$ forces the energy states onto distinct circles, determined by Equation \eqref{eq.vogelwiese}. (c) Constant density of states (DOS) of a 2DEG without any applied $B_z$. (d)-(f) Evolution of the DOS shown in (c) for increasing $B_z$. The DOS splits into LLs that are filled up to $E_\textrm{F}$ and are separated by $\hbar\omega_c$. At higher $B_z$, $\nu$ the number of LLs below $E_\textrm{F}$, decreases [Equation \eqref{eq.Labalaba}] and the DOS per LL increases [Equation \eqref{eq.Leberkaas}], to conserve the total number of available states. For a real system, the ideally $\delta$-shaped peaks broaden due to impurities present in the crystal lattice. When the peaks overlap like in (d), the classical Hall effect is observed. If the peaks are sufficiently spaced, the QHE takes over, as shown in panels (e) and (f). Figure adapted from ref. \cite{Gross2012}.}
  \label{fig_LandauLevel}
\end{figure}

 In 1980 von Klitzing discovered that a quantisation of the Hall resistance can occur at low temperature, under high applied magnetic fields \cite{Klitzing1980} [see Figure. \ref{fig_QHE}(b)]. To explain this effect, we consider a conventional two-dimensional electron gas (2DEG), for which we neglect spin dependence to simplify its description. The states of the 2DEG are uniformly distributed in $k$-space [Figure \ref{fig_LandauLevel}(a)], and the corresponding density of states (DOS) $D_\textrm{2D}$ is constant and independent of the system energy $E$ [Figure \ref{fig_LandauLevel}(c)]. If a perpendicular magnetic field $B_z$ is applied to the 2DEG, the Lorentz force causes the electrons to perform orbital motions with a certain cyclotron frequency $\omega_\textrm{c}=\frac{qB_z}{m}$, where $m$ is the effective electron mass. Furthermore, the electron gains an additional contribution $-e\cdot\vec{A}$ to its momentum, due to interactions with the electromagnetic field. 
\noindent The Hamiltonian describing a system of non-interacting charged particles in a magnetic field is then given by
\begin{align}
\label{eq.fleissigfleissig}
    \hat{H}=\frac{1}{2m}[\vec{p}-q\cdot\vec{A}(\vec{r})]^2.
\end{align}
The vector potential $\vec{A}(\vec{r})$ associated with the magnetic field $B_z$ has gauge freedom. If the Landau gauge is chosen such that $A_x=A_z=0$ and $A_y=x\cdot B_z$, the Hamiltonian given by Equation \eqref{eq.fleissigfleissig} reads
\begin{align}
\label{eq.HamiltonianHarmOszi}
    \hat{H}=\frac{1}{2m}\left[p_x^2+(p_y-qB_zx)^2\right]=\frac{1}{2m}\left[-\hbar^2\frac{\partial^2}{\partial x^2}+(-i\hbar\frac{\partial}{\partial y}+qB_zx)^2\right].
\end{align}
Since $\vec{A}(\vec{r})$ does not depend on $y$, we have translational invariance along the $y$-direction. As a result, the solution of the stationary Schrödinger equation is given by the product of a plane wave travelling along the $y$-direction with some yet-to-be-determined function in $x$-direction:
\begin{align}
\label{eq.AnsatzXYZ}
    \psi(x,y)=e^{ik_yy}\cdot u(x).
\end{align}
With the ansatz given by Equation \eqref{eq.AnsatzXYZ}, the stationary Schrödinger equation reads:
\begin{align}
\label{eq.SG_HarmOszi}
    \frac{1}{2m}\left(-\hbar^2\frac{\partial^2}{\partial x^2}+(\hbar q+eB_zx)^2\right)\cdot u(x)=Eu(x).
\end{align}

\noindent Equation \eqref{eq.SG_HarmOszi} describes the problem of a quantum harmonic oscillator, the solution of which yields quantised eigenenergies with the following form:\footnote{This is a well-known problem. An in-depth description of the problem of the quantum harmonic oscillator and of the derivation of its solutions can be found, e.g., in refs. \cite{Demtröder2016, Griffiths2012}.}
\begin{align}
\label{eq.vogelwiese}
    E_n=\left(n+\frac{1}{2}\right)\hbar\omega_\textrm{c},\quad \text{with }n\in\mathbb{N}_0.
\end{align}
Due to the quantisation of the eigenenergies \eqref{eq.vogelwiese}, the states populating the $k$-space are not distributed uniformly anymore but forced on distinct circles with radius $|k|=\sqrt{2E_nm/\hbar^2}$ [Figure \ref{fig_LandauLevel}(b)]. Also the DOS, which is constant in the absence of an applied $B_z$ or for small $B_z$, now splits up into "Landau levels" (LL). The LLs can be seen as a series of $\delta$ functions separated in energy by $\hbar\omega_\textrm{c}$, and each occupied by a large number of degenerate states [compare Figures \ref{fig_LandauLevel}(c)-(f)]. In real systems, the $\delta$ functions are broadened due to impurities in the crystal structure. The DOS $D_{\textrm{2D},B}$ is given by the ratio between the density $B_z$ of the total magnetic flux passing through the sample and the flux quantum $\phi_0=h/e$, which is the magnetic flux associated with a single state \cite{Gross2012}:
\begin{align}
\label{eq.Leberkaas}
    D_\textrm{2D,B}=\frac{eB_z}{h}.
\end{align}

\noindent In the ground state, the LLs are filled up to the Fermi energy $E_\textrm{F}$. If the applied field $B_z$ is increased, the distance $\hbar\omega_\textrm{c}$ between the levels also increases, and more energy levels are pushed above $E_\textrm{F}$ (Equation \eqref{eq.vogelwiese}). Since the total number of states is conserved, the DOS in magnetic field, $D_{\textrm{2D},B}$ must increase to compensate for the decrease in number of LLs that are below $E_\textrm{F}$ and can therefore be occupied (see Equation \eqref{eq.Leberkaas}). This behaviour is illustrated in Figure \ref{fig_LandauLevel}(d)-(f) for a progressively increasing $B_z$. The filling factor $\nu$, defined as the number of occupied LLs, is given by the ratio between the total number of states to the number of states per LL, which corresponds to the ratio of the DOS without and with applied magnetic field, meaning that
\begin{align}
\label{eq.Labalaba}
    \nu=\frac{D_\textrm{{2D}}}{D_{\textrm{2D},B}}=\frac{D_\textrm{2D}h}{eB_z}.
\end{align}
By solving Equation \eqref{eq.Labalaba} for $D_\textrm{2D}$, we can rewrite the expression for the Hall resistivity 
\eqref{eq.ClassicHallResistivity} as
\begin{align}
\label{eq.QuantizedHallResistivity}
    \rho_{yx}=\frac{B_z}{eD_\textrm{2D}}=\frac{h}{\nu e^2}.
\end{align}
Using Equation \eqref{eq.QuantizedHallResistivity}, the experimentally-observed behaviour of  $\rho_{yx}$ for increasing $B_z$ shown in Figure \ref{fig_QHE}(b) can be explained. For increasing $B_z$, the spacing of the LLs increases, and they cross $E_\textrm{F}$ one after the other. When the LLs overcome $E_\textrm{F}$, they depopulate [Figure \ref{fig_LandauLevel}(d)-(f)]. When $E_\textrm{F}$ falls within the gap between two LLs, $\nu$ takes an integer value and $\rho_{yx}$ exhibits a plateau, based on Equation \eqref{eq.QuantizedHallResistivity}. Whenever $E_\textrm{F}$ intersects with a LL, $\nu$ and in turn also $\rho_{yx}$ vary continuously until the next plateau is reached. This trend continues, as $B_z$ is increased, until only one LL is left below $E_\textrm{F}$ ($\nu=1$). In this situation, the lowest LL contains all available states, and the system reaches the so-called "magnetic quantum limit".

\par So far, we have discussed the simplified case for a 2DEG. However, in a real system other effects must be also considered. The Zeeman effect \cite{ZEEMAN1897} and the Paschen-Back effect \cite{Paschen1912}, for example, can induce a more complex splitting of the energy levels, which results in additional and irregularly spaced Hall plateaus in the resistance versus field $R$($B_z$) characteristics \cite{Movva2017, Kim2017}. Furthermore, the effect of electron scattering at impurities must be considered in the crystal lattice of a real system. According to the Drude model, electron scattering occurs on average at a time difference $\tau$. The energy-time uncertainty principle states that, for a state with an undisturbed lifetime $\tau$, its energy can be determined with an uncertainty $\Delta E\sim\hbar/\tau$. In turn, this relation leads to a broadening of the ideally $\delta$-shaped LLs described above. To observe the QHE this energy smearing must be smaller than the spacing $\hbar\omega_\textrm{c}$ of the LLs:
\begin{align}
    \Delta E\ll\hbar\omega_\textrm{c}\quad\longrightarrow\quad\frac{1}{\tau}\ll\frac{q}{m}B_z.
\end{align}
As a result, large magnetic fields $B_z$ are required to increase $\omega_\textrm{c}$, and measurements must be also carried out at low temperature to increase $\tau$. If these conditions are not fulfilled, the LLs overlap and the classical (linear) Hall effect is observed (Figure \ref{fig_QHE}(b) for $B_z < 1$\,T).\par
Since irregularities in the crystal structure occur randomly, it is difficult to describe them mathematically. One possible approximation to model their effects is by perturbing the flat LLs with a periodic potential. This approach has been followed by Thouless in 1982 \cite{Thouless1982}. Starting from the Kubo formula \cite{Kubo1957}, Thouless developed an expression for the Hall conductivity $\sigma_{xy}$ which is nowadays known as the TKNN-formula (named after Thouless, Kohmoto, Nightingale and Nijs) \cite{Thouless1982}:
\begin{align}
\label{eq.TKNN}
    \sigma_{xy}&=\frac{e^2}{2\pi h}\sum_n\int_{\text{BZ}}\text{d}^2k\underbrace{i\int_{\text{UC}}\text{d}^2r\left(\frac{\partial u_n^*}{\partial k_x}\frac{\partial u_n}{\partial k_y}-\frac{\partial u_n^*}{\partial k_y}\frac{\partial u_n}{\partial k_x}\right)}_{\substack{=i\bra{\nabla u_n}\times\ket{\nabla u_n}\stackrel{\text{\eqref{eq.Ohio}}}{=}\vec{\Omega}_n}}\\
    \label{eq.TKNN2}
    &\stackrel{\text{\eqref{eq.ChernNumber}}}{=}\frac{e^2i}{h}\sum_nQ_n,
\end{align}
where $u_n$ are Bloch eigenfunctions (Equation \eqref{eq.BlochsTheorem}) of the Hamiltonian given by Equation \eqref{eq.HamiltonianHarmOszi}. In Equation \eqref{eq.TKNN}, the sum is taken over the $n$ fully-occupied bands laying below $E_\text{F}$ and the integrals are taken over the first BZ and unit cell (UC). Using Equation \eqref{eq.Ohio} derived in the \hyperref[section-addendum]{Appendix}, it can be seen that the integral over the UC corresponds to the Berry curvature $\vec{\Omega}_n$ of the $n$-th band. We can therefore substitute the right-hand side of Equation \eqref{eq.TKNN} with the expression for the Chern number $Q_n$ given by Equation \eqref{eq.ChernNumber}.\footnote{In modern scientific literature, the Berry connection is defined as $\vec{A}=i\braket{n|\nabla n}$ \cite{Xiao2010}. However, in the context of the QHE, often the definition $\vec{A}=\braket{n|\nabla n}$ is used instead \cite{Kohmoto1985}. From the latter definition, one finds the better-known solution $\sigma_{xy}=\frac{e^2}{h}\sum_nQ_n$.} Comparison of Equations \eqref{eq.TKNN2} and \eqref{eq.QuantizedHallResistivity} shows that the filling factor $\nu$ is related to the sum of the Chern numbers of the occupied bands:
\begin{align}
\label{eq.Endspurt}
    \nu=i\sum_nQ_n.
\end{align}
Therefore, the origin of quantised Hall plateaus lies in the Berry curvature of the system, which is in turn related to the intrinsic properties of the band structure of the system.\par
As shown by Figure \ref{fig_QHE}(b), also the longitudinal resistivity $\rho_{xx}$ of a quantum Hall system shows an interesting behaviour. When $\rho_{yx}$ exhibits a plateau, $\rho_{xx}$  vanishes (i.e., $\rho_{xx}=0$), while $\rho_{xx}$ spikes during the transition of $\rho_{yx}$ between two Hall plateaus. This is the so-called Shubnikov-de Haas effect \cite{SCHUBNIKOW1930}. 
From Equation \eqref{eq.NougatBits} we also see that:
\begin{align}
\label{eq.rhoyx}
\forall \rho_{yx} \neq 0: \qquad \rho_{xx}=0 \quad \xrightarrow{\eqref{eq.NougatBits}} \quad \sigma_{xx}=0
\end{align}
This equation implies that we have a system that is simultaneously a perfect conductor and an insulator, when $\rho_{yx}$ reveals a plateau. This result appears counter-intuitive, but can be explained by deformations in the potential landscape of the system, which lead to dissipationless edge states. The following explains the mechanism underlying these edge states in more detail.
\par The potential which is associated with the energy of a LL is called Landau potential. In our previous elaborations, the Landau potential was flat. However, in a real sample the infinitely high potential of the surrounding vacuum causes a steep upward bending of the otherwise flat Landau potential close to the sample edges. There the Landau potential crosses the Fermi energy $E_\textrm{F}$. As a result, one-dimensional (1D), conducting channels form along the edges of the sample, each carrying a conductance of $e^2/h$, also known as conductance quantum\footnote{This is the value of the conductance quantum per spin direction.}. The total number of conducting channels corresponds to the number of LLs crossing $E_\textrm{F}$, which is the same as the filling factor $\nu$. Therefore, the transverse conductivity can be expressed as
\begin{align}
    \sigma_{xy}=\frac{\nu e^2}{h},
\end{align}
which is the inverse of the quantised Hall resistivity (Equation \eqref{eq.QuantizedHallResistivity}) derived earlier.\par
\begin{figure}[H]
  \centering
  \includegraphics[width=0.9\textwidth]{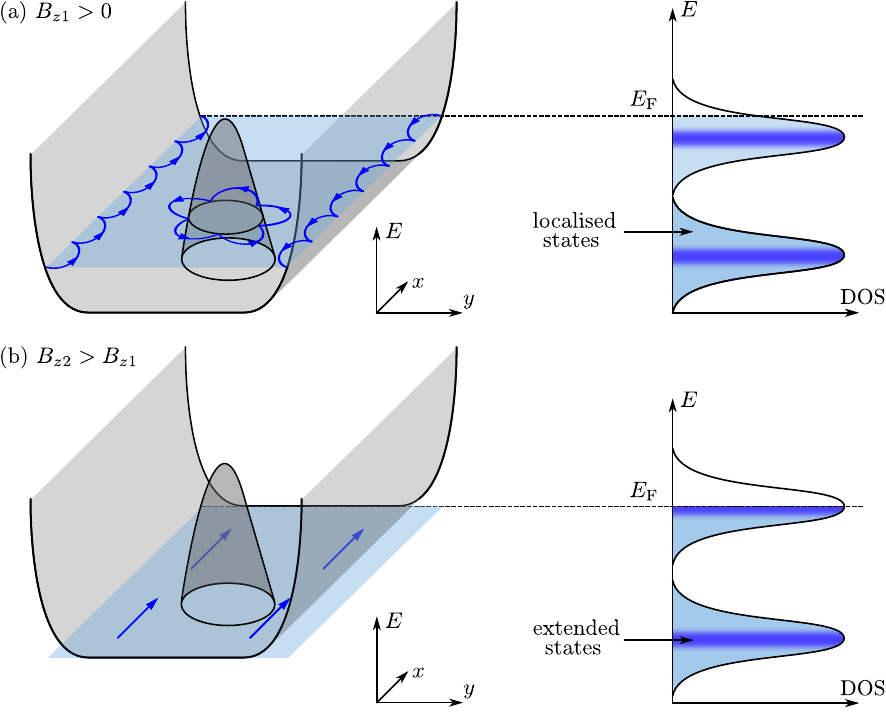}
  \caption[Localised states and edge states of a two-dimensional electron gas]{Localised and edge states of a 2DEG. Left: Potential landscape of a single LL. The flat potential is distorted by impurities (shown as a hump) and sharply bends upwards at the sample boundaries. Right: Corresponding density of states. (a) The centre of the LL lies below $E_\textrm{F}$. In this case, conducting channels form along the intersections at the sample edges and around impurities. Due to the applied magnetic field $B_z$, electrons move along the so-called "skipping orbits." The paths around the impurities are self-contained, so that the states are localised and do not contribute to the net charge transport. (b) As $B_z$ is further increased, the centre of the LL crosses $E_\textrm{F}$. In this situation, the conducting channels extend over the whole sample area and normal conducting behaviour is observed. Adapted from ref. \cite{Gross2012}.}
  \label{fig_EdgeStates}
\end{figure}

Due to the applied magnetic field $B_z$, all electrons perform orbital motions. Along the edges of the sample, the (otherwise closed) electron orbits are interrupted by the boundaries of the system, and their transversal momentum component (relative to the boundary) is reflected at the system edges, while the tangential one pertains. As a result, the current in the edge channels follows the so-called "skipping-orbit mechanism", as illustrated in Figure \ref{fig_EdgeStates}. The currents along opposite edges have opposite directions of propagation, such that the total current passing through the sample is zero ($\sigma_{xx}=0$). When an electron encounters an impurity (a potential barrier that acts as boundary), the magnetic field $B_z$ suppresses a backscattering of the electron via the same skipping orbit mechanism. Furthermore, scattering between charge carriers moving along different directions is prevented due to the spatial separation of the edge currents. Therefore, charge transport takes place without any dissipation ($\rho_{xx}=0$) and the potential at any point of an edge channel corresponds to the potential of the point where it originates. Consequently, if an external electric field $E_x$ is applied between the left and right edges of the sample, a Hall voltage proportional to $E_x$ can be measured between any couple of points at the top and bottom edges of the sample.

\indent Impurities in the crystal lattice also distort the potential landscape of the LLs, as shown by Figure \ref{fig_EdgeStates}(a), where an impurity is sketched as a hump in the potential. When this hump intersects $E_\textrm{F}$, an additional conducting channel forms, which allows electrons to run through skipping orbits along the intersection. If the impurity is located close to the sample edge, the intersection becomes part of the edge channel. If the impurity is instead located well inside the sample area, the intersection with $E_\textrm{F}$ is self-contained and the electrons are localised around the impurity and do not contribute to the overall charge transport. The enclosed area is called a quantum Hall droplet. The flanks of the DOS of the LLs are only populated by such localised states. Therefore, only if the centre of a LL crosses $E_\textrm{F}$, states that extend over the entire sample are obtained, and the system behaves like a regular conductor. This model explains the sharp transitions between plateaus as well as the extensive flatness of the Hall plateaus in Figure \ref{fig_QHE}(b).
\noindent Further information about the QHE and its connection to topology can be found, e.g., in refs. \cite{Bohm2013, Gross2012, Tong2016}.

\section{Anomalous and Spin Hall Effect}
\label{Section_AHEundSHE}
Topology is not only responsible for the quantisation of the classical linear Hall effect as described so far. Topological effects can in fact generate additional Hall effects, even in the absence of a magnetic field. Two of these effects are the anomalous and the spin Hall effect (AHE and SHE, respectively). In literature, they are often described as two different effects, but they can also be seen as two different manifestations of the same effect. In other terms, the AHE and SHE share the same origin but manifest themselves in different ways. While Hall discovered the AHE already in 1881 \cite{Hall1881}, it took more than a century until the SHE was experimentally observed for the first time \cite{Kato2004, Wunderlich2005}. We briefly review here how the theoretical understanding of the AHE developed, and then introduce the SHE in the context of the model of the AHE.\par
\begin{figure}[h]
  \centering
  \includegraphics[width=0.9\textwidth]{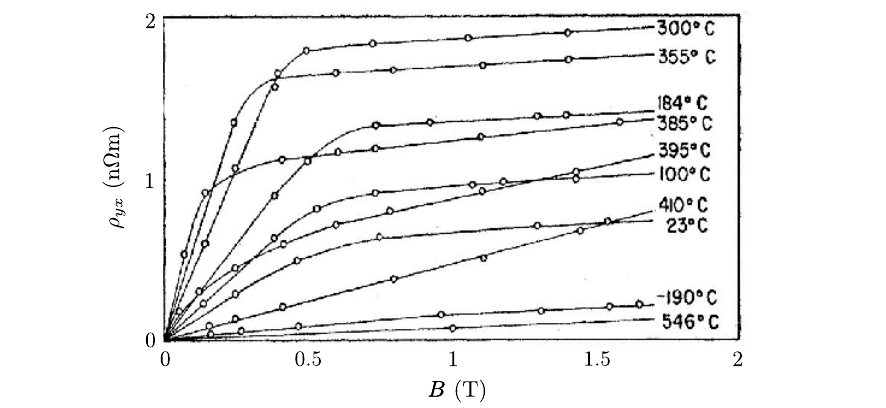}
  \caption[Anomalous Hall effect in Ni]{Anomalous Hall effect in Ni. The Hall resistivity $\rho_{yx}$ shows an additional component proportional to the magnetisation $M$ of the sample. When the maximum magnetisation is reached, this contribution remains constant, and the transport behaviour is then determined by the classical (linear) Hall effect. Picture adapted from ref. \cite{Pugh1953}.}
  \label{fig_AHE}
\end{figure}

Not only a magnetic field can cause a separation of charges transverse to an electric current, also a variety of other effects can contribute to the appearance of a Hall effect. Hall himself already reported the observation of an anomalous behaviour in the Hall resistivity curves measured for the ferromagnetic materials nickel (Ni) and cobalt  (Co) \cite{Hall1881}. A couple of years later, Smith studied the Hall resistivity of Ni in more detail \cite{Smith1910} reporting the results shown in Figure \ref{fig_AHE}. There, the Hall resistivity $\rho_{yx}$ exhibit a linearly-increasing behaviour as the applied $B_z$ is increased, which is then followed by a sharp bend, when the sample reaches its saturation magnetisation. After the bend, the curves increase again linearly, but more slowly. The conclusion is that, in addition to the classic Hall effect, a second mechanism causes a separation of charges transversal to the current direction and proportional to the magnetisation $M$ of the sample. Once $M$ saturates, the contribution of this anomalous effect is constant, and the additional increase in $\rho_{yx}$ with $B_z$ is only caused by the classical Hall effect. To describe this behaviour, a phenomenological term that is proportional to the magnetisation $M$ of the system was added to the expression for the classical Hall resistivity Equation \eqref{eq.ClassicHallResistivity}:
\begin{align}
    \rho_{yx}=\underbrace{A_\textrm{H}(T)\cdot B_z}_{\text{classical}}+\underbrace{A_\textrm{S}(T)\cdot M(T,B_z)}_{\text{anomalous}},
\end{align}
where $A_\textrm{S}$ is the spontaneous or also called extraordinary Hall constant, $T$ is the temperature.
\noindent The first microscopic theories on the AHE were developed in the 1950s \cite{Luttinger1954,Smit1955,Smit1958}, and nowadays three different mechanisms have been identified that can contribute to the AHE.\par
\begin{figure}[H]
  \centering
  \includegraphics[width=0.9\textwidth]{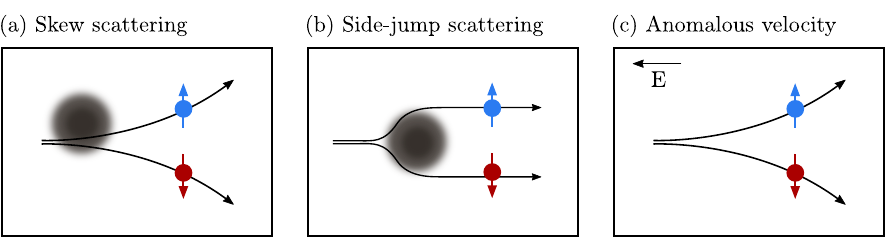}
  \caption[Contributions to the anomalous Hall effect]{Contributions to the anomalous Hall effect. (a) Skew scattering related to electrons experiencing a spin-dependent deflection due to scattering at an impurity. (b) Side-jump scattering related to electrons undergoing spin-dependent transverse displacement due to their interaction with an impurity or with a regular crystal potential. (c) Anomalous velocity related to electrons gaining a spin-dependent anomalous velocity component perpendicular to $\vec{E}$, whilst passing through a spin-orbit coupled crystal.}
  \label{fig_AHE_Contributions}
\end{figure}

The first contribution is an elastic, spin-dependent scattering of the electrons at impurities, known as skew or Mott scattering \cite{Smit1958}. Mott scattering is comparable to the classic Rutherford scattering, but the scattering cross section gains an additional spin-dependent component, which results in different direction-dependent scattering probabilities for spin-up and spin-down electrons [see Figure \ref{fig_AHE_Contributions}(a)]. Since ferromagnetic materials have an imbalance between the spin-up and spin-down electron populations, skew-scattering leads to a transversal separation of charges and consequently to a measurable Hall voltage. Skew-scattering only appears in disordered crystals, so it can be treated as an extrinsic contribution to the AHE.

\par The second contribution to the AHE is side-jump scattering, in which an electron receives a spin-dependent transverse displacement [Figure \ref{fig_AHE_Contributions}(b)]. This can happen when an electron is scattered at an impurity (extrinsic) or by interaction of an electron with the regular crystal potential (intrinsic). Berger observed that the density of impurities drops out when calculating the side-jump contribution to the AHE \cite{Berger1970}. Therefore, both extrinsic and intrinsic side-jump scattering can be treated as an intrinsic contribution to the AHE.\par
The third contribution to the AHE is given by the so-called anomalous velocity, introduced by Karplus and Luttinger in 1954 \cite{Luttinger1954}. The anomalous velocity is a purely topological effect that is closely related to the Berry curvature, and can therefore be seen as another intrinsic contribution.
\noindent Depending on the properties of a physical system, these different contributions to the AHE can be more or less pronounced. Overall, each mechanism independently contributes to the total transverse current. The observed Hall voltage is a result of the sum of these transverse currents. Because the conductivity is directly related to the current density ($J=\sigma E$), the total anomalous Hall conductivity $\sigma_{xy}^{\text{AH}}$ is the sum of the conductivities from each mechanism.
\begin{align}
    \sigma_{xy}^{\text{AH}}=\sigma_{xy}^{\text{skew}}+\sigma_{xy}^{\text{sj}}+\sigma_{xy}^{\text{av}}.
\end{align}\par
 In the following, we take a closer look at the anomalous velocity and at its connection with topology. The origin of the anomalous velocity can be found in the spin-orbit interaction of the electron's magnetic moment with the periodic crystal potential $V(\vec{r})=V(\vec{r}+\vec{e}_i)$ under the perturbation of a constant weak electric field $\vec{E}=\nabla\phi-\frac{\partial}{\partial\text{t}}\vec{A}$. The electric potential $\phi(\vec{r})$ destroys the translational invariance of the system and therefore Bloch's theorem, Equation \eqref{eq.BlochsTheorem}, does not hold. To circumvent this problem, we assume $\phi(\vec{r})=0$ and choose a time-dependent but spatially constant vector potential $\vec{A}(t)$ such that $-\frac{\partial}{\partial t} \vec{A}(t)=\vec{E}$. This method is called Peierls' substitution \cite{Peierls1933, Hofstadter1976}. The Hamiltonian of the system then reads
\begin{align}
\label{eq.nichtmehrsoheiss}
    \hat{H}=\underbrace{\frac{\left(p+e\vec{A}(t)\right)^2}{2m}-eV(\vec{r})}_{\substack{=\hat{H}_0}}-
    \underbrace{\frac{e\hbar}{4m^2c^2}(\nabla V(\vec{r})\times\vec{p})\cdot\vec{\sigma}}_{\substack{=\hat{H}_1}},
\end{align}
where $\hat{H}_0$ is the so-called Hofstadter Hamiltonian \cite{Hofstadter1976} and $\hat{H}_1$ describes the influence of spin-orbit interaction.\footnote{A detailed derivation of $H_1$ can be found in Chapter 5.3.4 of ref. \cite{Nolting2014}} 
Since $\vec{A}(t)$ is spatially constant in this representation, the wave functions can be again written as Bloch waves (Equation \eqref{eq.BlochsTheorem}) and, by using first-order perturbation theory, we can derive an expression for the group velocity $\vec{v}_n$ of a given state. The exact derivation is laborious and is shown in the \hyperref[section-addendum]{Appendix}, but its final result is that
\begin{align}
    \vec{v}_n=\vec{v}_0+\frac{e}{\hbar}\vec{E}\times\vec{\Omega}_n,
\label{eq.AnomalousVelocity}
\end{align}
where $\vec{v}_0=\partial\omega_n/\partial \vec{k}$ is the usual group velocity with which the envelope of the wave is travelling. However, due to the spin-orbit interaction with the periodic crystal potential, an additional component perpendicular to the applied electric field $\vec{E}$ and to the Berry curvature $\vec{\Omega}_n$ is obtained, which is given by the second term on the right-hand side of Equation \eqref{eq.AnomalousVelocity}. This term is the above-mentioned anomalous velocity. We also note that Equation \eqref{eq.AnomalousVelocity} has a form similar to that of the semiclassical equation of motion for a charged particle:
\begin{align}
    \hbar\dot{\vec{k}}=q\vec{E}+q\cdot\dot{\vec{r}}\times\vec{B}.
\label{eq.bitconeeect}
\end{align}
\noindent In analogy to Equation \eqref{eq.bitconeeect}, one can imagine the term related to the anomalous velocity in Equation \eqref{eq.AnomalousVelocity} as a generalised Lorentz force defined in $k$-space.
\begin{figure}[H]
  \centering
  \includegraphics[width=0.9\textwidth]{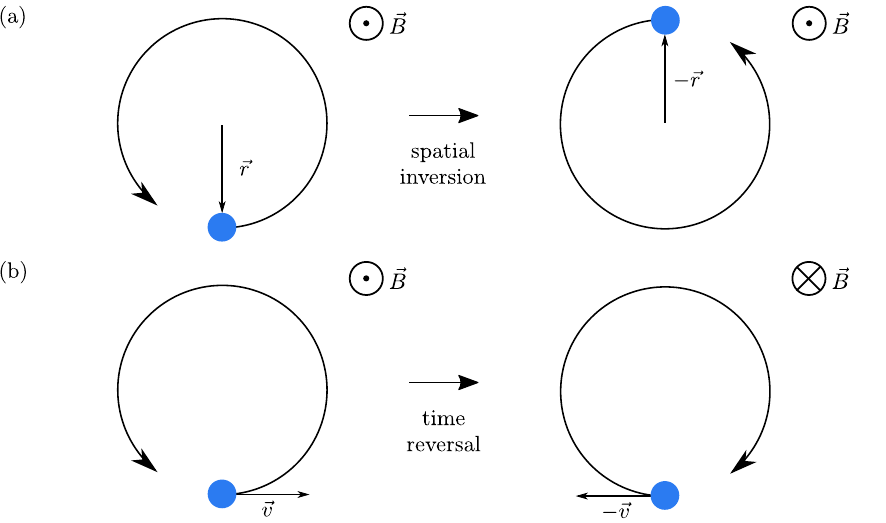}
  \caption[Spatial inversion and time reversal symmetry]{Spatial inversion and time reversal of an electron in a magnetic field $\vec{B}$. Under spatial inversion (a), the position coordinate is inverted, i.e., $\vec{r}\rightarrow -\vec{r}$, whilst the magnetic field is preserved which implies that $\vec{B}\rightarrow\vec{B}$. (b) Under time reversal, the velocity is inverted,  i.e., $\vec{v}\rightarrow -\vec{v}$ and therefore, to fulfil inversion symmetry, also the magnetic field must change sign which implies that $\vec{B}\rightarrow-\vec{B}$.}
  \label{fig_AHE_Symmetry}
\end{figure}
\par We now analyse how the anomalous velocity is related to the symmetry of a system. If the unperturbed system (i.e., without the weak electric field) preserves time-reversal and spatial-inversion symmetry, then also Equation \eqref{eq.AnomalousVelocity} must be invariant under such transformations. Figure \ref{fig_AHE_Symmetry} shows the spatial- and time-inversion for an electron moving in a magnetic field $\vec{B}$.
\noindent We see that the magnetic field $\vec{B}$ is invariant under spatial inversion but changes sign under time reversal. The same conditions apply to the Berry curvature in $k$-space:
\begin{align}
    \vec{\Omega}_n\xrightarrow{\text{s.i.}}\vec{\Omega}_n\quad\text{and}\quad  \vec{\Omega}_n\xrightarrow{\text{t.r.}}-\vec{\Omega}_n.
\end{align}
As a result, we infer that, for systems that preserve both spatial inversion and time reversal symmetries, the Berry curvature $\Omega_n$ must be zero, and Equation \eqref{eq.AnomalousVelocity} reduces to the conventional expression for the group velocity. Therefore, the anomalous velocity can only affect systems with either broken spatial-inversion symmetry or broken time-reversal symmetry.\par
\begin{figure}[H]
  \centering
  \includegraphics[width=0.9\textwidth]{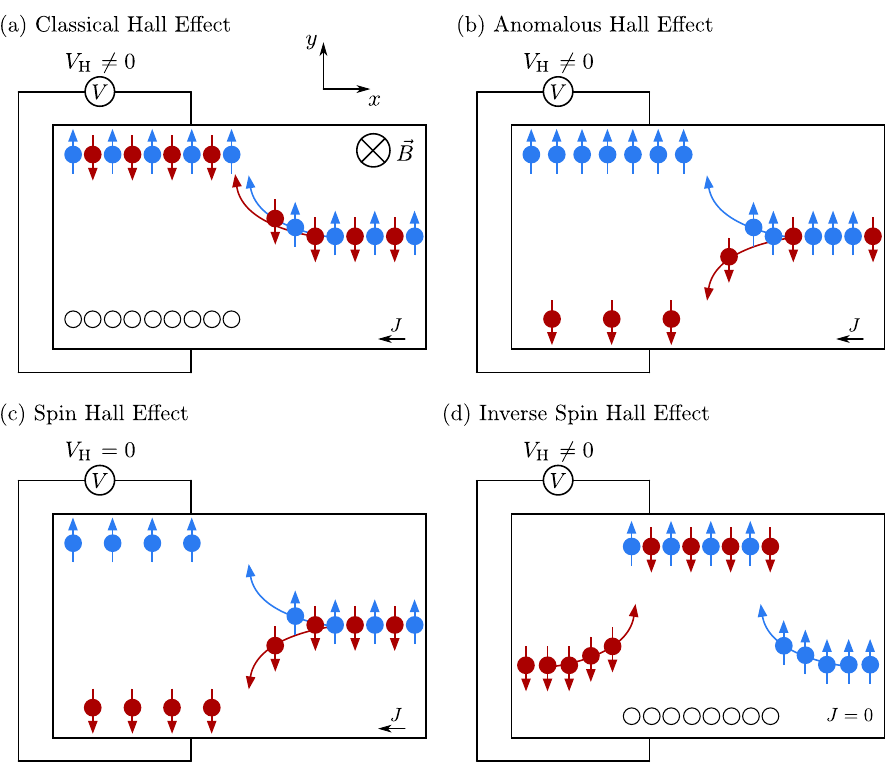}
  \caption[Overview over the different Hall effects.]{Overview over the different Hall effects for spin-up ($\uparrow$) and spin-down ($\downarrow$) electrons. (a) Classical Hall effect: a current is deflected due to an applied magnetic field $B_z$. This effect is spin independent. (b) Anomalous Hall effect (AHE): $\uparrow$ and $\downarrow$ electrons are deflected in opposite directions. In systems with spin imbalance, this causes a separation of charges leading to the appearance of a Hall voltage $V_\textrm{H}$. (c) Spin Hall effect (SHE): in systems with equal numbers of $\uparrow$ and $\downarrow$ electrons, two separated spin currents form and $V_\textrm{H}=0$. (d) Inverse spin Hall effect (ISHE): in systems with a pure spin current, $\uparrow$\,\,and $\downarrow$ electrons are deflected in the same direction leading to a Hall voltage $V_\textrm{H}$.} 
  \label{fig_Hall_Overview}
\end{figure}

Let us consider now a system where one of the symmetries (time reversal or spatial inversion) is broken and $\Omega_n\neq0$ holds. In this system, the anomalous velocity \eqref{eq.AnomalousVelocity} causes a deflection of the charge carriers perpendicular to the Berry curvature $\Omega_n$ and to the applied electric field $\vec{E}$, which gives rise to a Hall effect. The sign of $\Omega_n$ is spin-dependent \cite{Berry1984}. Therefore, spin-up and spin-down electrons are deflected in opposite directions. Depending on the spin composition of the current, three different kinds of Hall effects can manifest: the AHE, the SHE, and the inverse spin Hall effect (ISHE), which are illustrated in Figure \ref{fig_Hall_Overview}. Unlike the classic and the QHE, the AHE, SHE and ISHE do not require the presence of a magnetic field.

\par The AHE occurs in systems with broken time-reversal symmetry. Such systems, like ferromagnetic metals (e.g., Co, Ni etc.), have an unequal number of spin-up and spin-down electrons. The three contributions to the AHE explained earlier (i.e., skew scattering, side-jump scattering, and anomalous velocity) cause a deflection of the spin-up and spin-down electrons in opposite directions, perpendicular to the direction of their transport. Due to the spin imbalance, this leads to a separation of charges and to a measurable Hall voltage $V_\textrm{H}$ [Figure \ref{fig_Hall_Overview}(b)].
\par The SHE occurs in systems that do not break time-reversal symmetry because they have an equal number of spin-up and spin-down electrons. In these systems, the same number of electrons (with oppositely-oriented spins) will be deflected along opposite directions perpendicular to their motion. In this case, $V_\textrm{H}=0$, but the system exhibits two separated spin currents along its edges [Figure \ref{fig_Hall_Overview}(c)].
\par The ISHE occurs under flow of pure spin currents. Pure spin current means that the same number of spin-up and spin-down electrons are running in opposite directions. In this case, only spins are transported and the net charge current is zero. In other words, there are two spin currents that have oppositely-oriented spins and that are flowing in opposite directions. Therefore, the spin-up and spin-down electrons are deflected towards the same direction, leading to a separation of charges and a non-vanishing $V_\textrm{H}$ [Figure \ref{fig_Hall_Overview}(d)]. How a pure spin current can be generated in a system is explained, e.g., in refs. \cite{Takahashi2016, Adachi2013, Jedema2002, Valenzuela2004}.
\par In summary, we see that the AHE, SHE and ISHE share the same origin, namely the anomalous velocity, skew- and side-jump scattering, but they manifest in different ways. Depending on the initial composition of spin-up and spin-down electrons and their directions of travel, the deflection due to the anomalous velocity, skew- and side-jump scattering causes a separation of charges (ISHE) or spins (SHE) or both (AHE).
\par In general, the electron spin is not preserved in a material due to spin-flip scattering, which can be caused by the interactions of the electron spin with other electrons, phonons, or impurities. Therefore, the spin-dependent Hall effects described above typically manifest only in systems with a long spin-diffusion length $\lambda$. A material class that preserves the electron spin, and is therefore suitable for the observation of topological Hall effects, is that of topological insulators. They are the topic of the following chapter. Further information about the anomalous velocity and the different types of Hall effects discussed in this chapter can be found in refs. \cite{Luttinger1954, Xiao2010, Sinova2015, Nagaosa2010}.

\section{Topological Insulators}
\label{section_topologischeIsolatoren}
Topological insulators (TIs) are materials that are insulating in the bulk but exhibit topologically-protected dissipationless conducting edge states. A system like a TI has been already introduced in the context of the QHE [see Chapter \ref{section_QHE}]. There we have seen that a strong external magnetic field $B$ can suppress the conductivity in the bulk, and generate conducting channels (skipping orbits) along the edges. The total number of conducting channels is related to the Chern number, a topological invariant of the system [see Equation \eqref{eq.Endspurt}]. From a more general point of view, it turns out that not the applied $B$ but rather the nontrivial topology of the band structure of the system is the key ingredient to get robust conducting edge states. In 2006 Bernevig made the first theoretical prediction of the topological insulating properties of HgTe quantum wells in the absence of an applied $B$ \cite{Bernevig2006}, which was later experimentally confirmed by König et al. \cite{Koenig2007}. To date, a significant number of other systems have been demonstrated to be TIs \cite{Chadov2010, Eremeev2012, Ando2013, Bansil2016, Liu2019}.
\par To understand the mechanism that generates topologically-protected edge states, we consider a system with a band gap and a nontrivial Berry curvature. Following Equation \eqref{eq.ChernNumber}, the $n$-th band has a certain topologically-invariant Chern number $Q_n$. Under the assumption that the adiabatic approximation holds, the Hamiltonian and the underlying band structure of the system can be continuously deformed without changing the topological invariants of the system itself, as long as the bands are sufficiently separated and the gap around $E_\textrm{F}$ remains open. In turn, to change a topological invariant, the gap to $E_\textrm{F}$ must be closed. To meet both requirements, the system must exhibit a closed gap on the surface, where the nontrivial topological invariant of its band structure changes to the trivial value of the vacuum surrounding the system. This leads to conducting edge states in the case of two-dimensional (2D) TIs and to conducting surface states for three-dimensional (3D) TIs. A similar phenomenology can be observed in the QHE. Here, the bulk of a material is insulating when the system's Hall resistance corresponds to a plateau value. But as soon as the system switches from one plateau to another, which corresponds to a change of the topological invariant, the system must pass through a conductive phase, which results in peaks in its longitudinal resistance [compare Figure \ref{fig_QHE}].\par
\begin{figure}[H]
  \centering
   \includegraphics[width=0.9\textwidth]{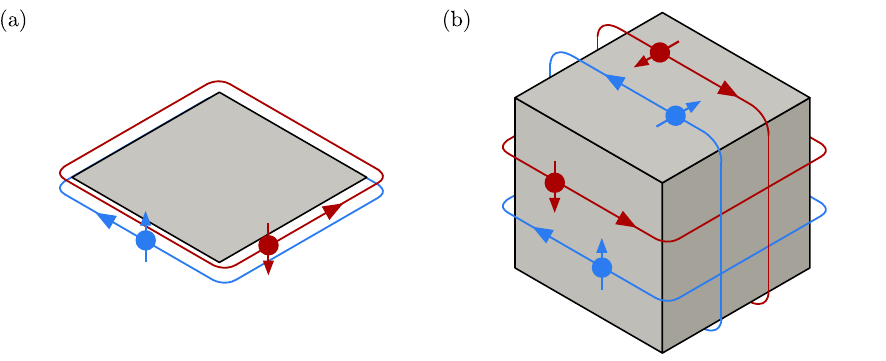}
  \caption[Edge and surface states of topological insulators.]{Helical edge and surface states of 2D (a) and 3D (b) topological insulators. Due to strong spin-orbit interaction, the spin is momentum-locked perpendicular to the momentum and to the normal vector of the surface. States moving along opposite directions have oppositely-oriented spins.} 
  \label{fig_SurfaceStates}
\end{figure}

TIs are systems with strong spin-orbit interaction \cite{Qi2011}. Recalling the Hamiltonian given by Equation \eqref{eq.nichtmehrsoheiss}, we see that the energy associated with spin-orbit interactions involves the cross product of the gradient of the crystal potential with respect to the momentum. The potential gradient at the boundary between the crystal and the vacuum is perpendicular to the surface, while the momentum vector of surface states lies within the surface. Consequently, the vector obtained from the cross product, to which the spin is momentum-locked, lies within the crystal surface and is perpendicular to the momentum. Therefore, states that are travelling in opposite directions must have spins with opposite orientations. These types of surface states are also referred to as helical edge states \cite{Wu2006} and they are illustrated for 2D and 3D TIs in Figure \ref{fig_SurfaceStates}.
\noindent When a helical edge state is scattered back at a nonmagnetic impurity, it changes its spin by either $\pi$ or $-\pi$ due to the spin-momentum locking. The two possible states differ by a rotation of $\theta=2\pi$. The corresponding rotation operator for spin-1/2 particles reads
\begin{align}
    e^{i\sigma\theta}=e^{i\frac{1}{2}2\pi}=-1.
\end{align}
We see that a wave function picks up a negative sign under a $2\pi$ rotation.\footnote{This explanation is strongly simplified. A more formal explanation can be found in ref. \cite{Streater2016}.} Consequently, the wave functions of backscattered states interfere destructively on average, which corresponds to a perfect transmission. However, if the impurity is magnetic (i.e., if time-reversal symmetry is broken), then the states can also backscatter by spin-flip scattering and do not necessarily interfere destructively \cite{Vaitkus2022}. Further information about topological insulators can be found in refs. \cite{Qi2011, Ortmann2015, Asboth2015}.

\section{Conclusion and Outlook}
Topology is a well-established branch of mathematics, focused on classifying objects based on the symmetry properties of their geometry. Although pioneering theoretical works, which applies topology to phenomena in solid state physics, have been published already in the 1960s, the potential of topology to explain, relate and predict physical phenomena has been understood only over last two decades. While it is now clear that fascinating observations such as, for example, the quantum Hall and anomalous Hall effects, are topological phenomena, topology has not yet become a standard part of most physics study programs, probably due to its abstract and formal nature. To close this gap, this tutorial introduces important concepts of topology, namely the Berry phase, the Berry curvature and the Chern number, and highlights their implications for solid-state physics. We have showed the relation between a non-trivial Berry phase and the topological properties of quantum systems. Specifically, we have elaborated the connection between the Berry phase and the band structure of a solid and the importance of cyclic evolutions, which are in turn related to the gauge invariance of the Berry curvature. We then have exemplified some relevant topological effects arising from the Berry phase like the quantum, anomalous, and spin Hall effects. We have then closed this tutorial by discussing the transport properties of topological insulators, which also originate from their Berry phase physics.
\par As a result of the strong connection between the Berry phase and the topological properties of quantum materials, several groups have recently started to explore different approaches to actively tune the Berry phase and to achieve control over related topological effects. The studied approaches for "Berry-phase engineering" include the design of non-collinear spin textures at material interfaces, which can be tuned via strain \cite{Groenendijk2020, Cuoco2022}, the application of gate voltage to ultrathin (2D) magnetic systems \cite{Du2020}, and the optical excitation of polariton systems \cite{Polimeno2021} or spin-based qubits \cite{Yale2016}.
We expect that in the near future these studies can lead to the development of devices for Berry electronics (Berrytronics), where information is encoded via a Berry-phase-induced change in topological states or effects, and also that a variety of other novel topological phenomena intrinsically connected with the Berry phase will be predicted or discovered. We therefore hope that this tutorial can contribute towards putting its readers in the position to actively advance these new fascinating research trends.

\newpage
\section*{Declarations}

\begin{itemize}
\item Conflict of interest
\newline The authors declare no conflict of interest.
\newline
\item Author contribution
\newline The manuscript is based on the Master Thesis of N.S.. A.D.B. supervised the Thesis project. A.D.B. and E.S. revised the manuscript.
\end{itemize}

\newpage

\begin{appendices}

\section{}
\label{section-addendum}
\subsection*{Berry Curvature as Sum Over the Eigenstates}
In the following, we show how to derive the formula for the Berry curvature as sum over the eigenstates (Equation \eqref{eq.crazyFormel}) from the initial expression (Equation \eqref{eq.BerryCurvature}). For a better overview we will not explicitly write out the $\vec{R}$ dependence of $\ket{n}, \hat{H}, E_n,\text{ and }\nabla$; furthermore $\nabla\ket{n}:=\ket{\nabla n}$. We start by rewriting Equation \eqref{eq.BerryCurvature} using the following correlation:
\begin{align}
\label{eq.HotMilk}
    \nabla\times\braket{A|\nabla A}&=\nabla\times
                                \int_{\mathbb{R}^3} \text{d}^3x\quad\braket{A|x}\nabla\braket{x|A}\notag\\
    &=\int_{\mathbb{R}^3}\text{d}^3x\quad\nabla\times(\underbrace{A^*}_{\substack{\text{scalar}}}
                                \underbrace{\nabla\cdot A}_{\substack{\text{vector}}}) \notag\\
    &=\int_{\mathbb{R}^3}\text{d}^3x\quad(\nabla A^*)\times(\nabla A)+A^*
                                \underbrace{\nabla\times(\nabla A)}_{\substack{=0}} \notag\\
    &=\bra{\nabla A}\times \ket{\nabla A}.
\end{align}
Step 3 of Equation \eqref{eq.HotMilk} is done by applying the product rule. How the product rule applies here becomes more obvious when looking at the curl in Einstein notation:
\begin{align}
    \nabla\times(\alpha\vec{A})=\epsilon_{ijk}\partial_j(\alpha\vec{A})=\epsilon_{ijk}(\partial_j\alpha)A_k+\epsilon_{ijk}\alpha(\partial_jA_k)=(\nabla\alpha)\times\vec{A}+\alpha\nabla\times\vec{A},
\end{align}
with $\epsilon_{ijk}$ being the Levi-Civita symbol. Using the correlation given by Equation \eqref{eq.HotMilk} and the expression for the Berry connection (Equation \eqref{eq.BerryVectorPotential}) we can rewrite the Berry curvature given by Equation \eqref{eq.BerryCurvature} as
\begin{align}
\label{eq.Ohio}
    \vec{\Omega}=\nabla\times\vec{A}=i\nabla\times\braket{n|\nabla n}=i\bra{\nabla n}\times\ket{\nabla n}.
\end{align}
Using the completeness of the basis $1=\sum_{m=0}^\infty \ket{m}\bra{m}$ we can rewrite
\begin{align}
\label{eq.hihihi}
    \bra{\nabla n}\times\ket{\nabla n}&=\bra{\nabla n}\left(\sum_m \ket{m}\bra{m}\right) \times \ket{\nabla n} \notag\\
    &=\sum_m \int_{\mathbb{R}^3}\text{d}^3x\quad\left((\nabla n^*)m\cdot m^*\right)\times(\nabla n) \notag\\
    &=\sum_m \int_{\mathbb{R}^3}\text{d}^3x\quad((\nabla n^*)m)\times (m^*\nabla n) \notag\\
    &=\sum_m \braket{\nabla n|m}\times\braket{m|\nabla n} \notag\\
    &=-\sum_m \braket{n|\nabla m}\times\braket{m|\nabla n}\notag\\
    &=\sum_{m\neq n} \braket{m|\nabla n}\times \braket{n|\nabla m}.
\end{align}
Step 5 of Equation \eqref{eq.hihihi} uses partial integration and the orthonormality of the basis:
\begin{align}
    \braket{\nabla n|m}=\int_{\mathbb{R}^3}\text{d}^3x\quad(\nabla n^*)m=\underbrace{\left[n^*\cdot m\right]_{\mathbb{R}^3}}_{\substack{=0}}-\int_{\mathbb{R}^3}\text{d}^3x\quad n^*\nabla m =-\braket{n|\nabla m}.
\end{align}
Step 6 of Equation \eqref{eq.hihihi} uses the antisymmetry of the cross product, allowing the exclusion of the case $m=n$.
\begin{align}
    \text{For }m=n:\qquad\braket{n|\nabla n}\times \braket{n|\nabla n}= - \braket{n|\nabla n}\times \braket{n|\nabla n}=0.
\end{align}
Using Equation \eqref{eq.hihihi} the Berry curvature Equation \eqref{eq.Ohio} can now be written as
\begin{align}
\label{eq.Kopfweh}
    \vec{\Omega}=i\sum_{m\neq n} \braket{m|\nabla n}\times \braket{n|\nabla m}.
\end{align}
To further evaluate the brackets in Equation \eqref{eq.Kopfweh}, we take the exterior derivative of the instantaneous eigenvalue equation (Equation \eqref{eq.InstantEigenvalueEquation}):
\begin{align}
\label{eq.tralala}
    \nabla\hat{H}\ket{n}+\hat{H}\ket{\nabla n}=\nabla E_n\ket{n}+E_n\ket{\nabla n}.
\end{align}
Multiplying Equation \eqref{eq.tralala} from left by $\bra{m}$ gives
\begin{align}
    \bra{m}\nabla\hat{H}\ket{n}+\bra{m}\hat{H}\ket{\nabla n}=\underbrace{\bra{m}\nabla E_n\ket{n}}_{\substack{=0}}+\bra{m}E_n\ket{\nabla n}.
\end{align}
Using linearity, normality (see Equation \eqref{eq.InstEigenstatesNormalized}), Equation \eqref{eq.InstantEigenvalueEquation} and solving for $\braket{m|\nabla n}$ we obtain
\begin{align}
\label{eq.Nablablabla}
    \braket{m|\nabla n}=\frac{\bra{m}\nabla\hat{H}\ket{n}}{E_n-E_m}.
\end{align}
Finally, by inserting Equation \eqref{eq.Nablablabla} into Equation \eqref{eq.Kopfweh} we find the expression for the Berry curvature as sum over the eigenstates of the equation \eqref{eq.crazyFormel}.
\begin{align}
     \vec{\Omega}(\vec{R})=i\sum_{m\neq n}\frac{\bra{n(\vec{R})}\nabla_{\vec{R}}\hat{H}(\vec{R})\ket{m(\vec{R})}\times\bra{m(\vec{R})}\nabla_{\vec{R}}\hat{H}(\vec{R})\ket{n(\vec{R})}}{\left(E_n(\vec{R})-E_m(\vec{R})\right)^2}. \tag{\ref{eq.crazyFormel}}
\end{align}

\newpage
\subsection*{Anomalous Velocity}
In the following, we show how to derive Equation \eqref{eq.AnomalousVelocity} for the anomalous velocity from the Hamiltonian \eqref{eq.nichtmehrsoheiss}. First, we need to derive an expression that will be important in the later elaborations. We start by differentiating the time-independent Schrödinger equation with respect to $\vec{k}$ and multiplying with $\bra{u_m}$ from the left:
\begin{align}
\label{eq.Doener}
    \bra{u_m}\frac{\partial\hat{H}}{\partial\vec{k}}\ket{u_n}+\bra{u_m}\hat{H}\ket{\frac{\partial u_n}{\partial\vec{k}}}=\bra{u_m}\frac{\partial E_n}{\partial\vec{k}}\ket{u_n}+E_n\braket{u_m|\frac{\partial u_n}{\partial\vec{k}}}.
\end{align}
Usually, the Hamiltonian $\hat{H}$ is Hermitian, so $\hat{H}=\hat{H}^\dag$, with $\hat{H}^\dag$ being the self-adjoint of $\hat{H}$. Hence, we can write $\bra{u_m}\frac{\partial\hat{H}}{\partial\vec{k}}\ket{u_n}$ as $ \left(\bra{u_n}\frac{\partial\hat{H}}{\partial\vec{k}}\ket{u_m}\right)^*$. Using this identity and subtracting the second term of the left-hand side, Equation \eqref{eq.Doener} becomes
\begin{align}
   \left(\bra{u_n}\frac{\partial\hat{H}}{\partial\vec{k}}\ket{u_m}\right)^*=\frac{\partial E_n}{\partial\vec{k}}\braket{u_m|u_n}+E_n\braket{u_m| \frac{\partial u_n}{\partial\vec{k}}}-\left(\bra{\frac{\partial u_n}{\partial\vec{k}}}\hat{H}\ket{u_m}\right)^*.
\end{align}
Taking the complex conjugate gives
\begin{align}
\label{eq.kalalada}
   \bra{u_n}\frac{\partial\hat{H}}{\partial\vec{k}}\ket{u_m}=
   \frac{\partial E_n}{\partial\vec{k}}\braket{u_n|u_m}
   +E_n\braket{\frac{\partial u_n}{\partial\vec{k}}|u_m}-\bra{\frac{\partial u_n}{\partial\vec{k}}}\hat{H}\ket{u_m}.
\end{align}
Using the orthonormality of $\ket{u_n}$ and $\hat{H}\ket{u_m}=E_m\ket{u_m}$, Equation \eqref{eq.kalalada} can be written as
\begin{align}
\label{eq.OffensichtlicherZusammenhang}
    m\neq n:\qquad \bra{u_n}\frac{\partial\hat{H}}{\partial\vec{k}}\ket{u_m}=\left(E_n-E_m\right)\braket{\frac{\partial u_n}{\partial\vec{k}}|u_m}.
\end{align}

\par Now we can start with the derivation of the anomalous velocity, which is based on the elaborations in \cite{Luttinger1954, Xiao2010}. As a reminder, the Hamiltonian of the system, that we got via Peierls substitution, reads
\begin{align}
    \hat{H}=\underbrace{\frac{\left(p+e\vec{A}(t)\right)^2}{2m}-eV(\vec{r})}_{\substack{=\hat{H}_0}}-
    \underbrace{\frac{e\hbar}{4m^2c^2}(\nabla V(\vec{r})\times\vec{p})\cdot\vec{\sigma}}_{\substack{=\hat{H}_1}}, \tag{\ref{eq.nichtmehrsoheiss}}
\end{align}
where $\hat{H}_1$ describes the influence of spin-orbit interaction. We can treat $H_1$ as a small perturbation. With a sufficiently small $\lambda\in\mathbb{R}$ first-order perturbation theory then gives
\begin{align}
\label{eq.esistimmernochheiss}
    \ket{\psi_n}=\ket{u_n^{(0)}}+\lambda\ket{u_n^{(1)}}=\ket{u_n^{(0)}}+\lambda\sum_{m\neq n}\ket{u_m^{(0)}}\frac{\bra{u_m^{(0)}}-\hat{H}_1\ket{{u_n^{(0)}}}}{E_n^{(0)}-E_m^{(0)}}.
\end{align}
\newpage 
\noindent With the time-dependent Schrödinger equation $i\hbar\frac{\partial}{\partial t} \ket{u_n}=\hat{H}_0\ket{u_n}+\lambda\hat{H}_1\ket{u_n}$ and $\braket{u_m|u_n}=\delta_{mn}$ we can rewrite
\begin{align}
    \bra{u_m^{(0)}}-\hat{H}_1\ket{{u_n^{(0)}}}&=\bra{u_m^{(0)}}-\frac{i\hbar}{\lambda}\partial t+\frac{\hat{H}_0}{\lambda}\ket{u_n^{(0)}}\\
    &=-\frac{i\hbar}{\lambda}\bra{u_m^{(0)}}\frac{\partial}{\partial t}\ket{u_n^{(0)}}+\frac{E_n^{(0)}}{\lambda}\underbrace{\braket{u_m^{(0)}|u_n^{(0)}}}_{\substack{=0}\text{ for }m\neq n}.
    \label{eq.esistheiss}
\end{align}
Inserting Equation \eqref{eq.esistheiss} into Equation \eqref{eq.esistimmernochheiss} gives the following expression for the wave function in first-order perturbation theory:
\begin{align}
    \ket{\psi_n}=\ket{u_n^{(0)}}-i\hbar\sum_{m\neq n}\frac{\ket{u_m^{(0)}}\bra{u_m^{(0)}}\frac{\partial}{\partial t}\ket{{u_n^{(0)}}}}{E_n^{(0)}-E_m^{(0)}}.
\end{align}
From now on we will write $\ket{u_n^{(0)}}$ as $\ket{u_n}$, etc. The group velocity of a wave function can be described by the velocity operator which in $k$-space reads $\hat{v}_n=\partial \hat{H}/(\hbar\partial\vec{k})$. We are interested in the average velocity of a given state $\psi_n$:
\begin{align}
    \bra{\psi_n}\hat{v}_n\ket{\psi_n}&=
    \left(\vphantom{\sum_{\neq}
    \frac{\frac{\partial}{\partial}}{E_n}}\right.
    \underbrace{\bra{u_n}\vphantom{\sum_{\neq}\frac{\frac{\partial}{\partial}}{E_n}}}_{\substack{:=a}}
    +\underbrace{i\hbar\sum_{m\neq n}
    \frac{\braket{\frac{\partial u_n}{\partial t}|u_m}\bra{u_m}}{E_n-E_m}}_{\substack{:=b}}  
    \left.\vphantom{\sum_{\neq}
    \frac{\frac{\partial}{\partial}}{E_n}}\right)
    \frac{\partial\hat{H}}{\hbar\partial\vec{k}}
    \left(\vphantom{\sum_{\neq}\frac{\frac{\partial}{\partial}}{E_n}}\right.
    \underbrace{\ket{u_n}\vphantom{\sum_{\neq}\frac{\frac{\partial}{\partial}}{E_n}}}_{\substack{=a^*}}
    -\underbrace{i\hbar\sum_{m\neq n}\frac{\ket{u_m}\braket{u_m|\frac{\partial u_n}{\partial t}}}{E_n-E_m}}_{\substack{=b^*}}
    \left.\vphantom{\sum_{\neq}
    \frac{\frac{\partial}{\partial}}{E_n}}\right)\\
    &=\frac{1}{\hbar}\left(\vphantom{\frac{\hat{H}}{k}}\right.
    \underbrace{a\frac{\partial\hat{H}}{\partial\vec{k}}a^*}_{=\partial E_n/\partial\vec{k}}-a\frac{\partial\hat{H}}{\partial\vec{k}}b^*+b\frac{\partial\hat{H}}{\partial\vec{k}}a^*-\underbrace{b\frac{\partial\hat{H}}{\partial\vec{k}}b^*}_{\stackrel{\text{second order}}{\approx\,\,0}}
    \left.\vphantom{\frac{\hat{H}}{k}}\right)\\
    &=\frac{1}{\hbar}\frac{\partial E_n}{\partial\vec{k}}
    -i\sum_{m\neq n}\frac{\bra{u_n}\frac{\partial\hat{H}}{\partial\vec{k}}\ket{u_m}\braket{u_m|\frac{\partial u_n}{\partial t}}-\braket{\frac{\partial u_n}{\partial t}|u_m}\bra{u_m}\frac{\partial\hat{H}}{\partial\vec{k}}\ket{u_n}}{E_n-E_m}.
    \label{eq.Muesli}
\end{align}
Applying Equation \eqref{eq.OffensichtlicherZusammenhang} and using the completeness of the basis $1=\sum_{m=0}^\infty \ket{u_m}\bra{u_m}$ we can rewrite Equation \eqref{eq.Muesli} as
\begin{align}
    \vec{v}_n=\bra{\psi_n}\hat{v}_n\ket{\psi_n}=\frac{1}{\hbar}\frac{\partial E_n}{\partial\vec{k}}
    -i\left(\braket{\frac{\partial u_n}{\partial\vec{k}}|\frac{\partial u_n}{\partial t}}
    -\braket{\frac{\partial u_n}{\partial t}|\frac{\partial u_n}{\partial\vec{k}}}\right).
\label{eq.gameboy}
\end{align}
We want to find a time-independent expression for $\vec{v}_n$. To this end, we use the semiclassical equation of motion for an electron with momentum $\vec{p}=\hbar\vec{k}$. It reads
\begin{align}
    \hbar\dot{\vec{k}}=-e\vec{\varepsilon}-e\cdot\dot{\vec{r}}\times\vec{B},
\end{align}
with the two contributions of the electrical and the Lorentz force. Notice that here we write the electric field as $\vec{\varepsilon}$ for a clear distinction from the energy eigenvalues $E_n$. However, in Chapter \ref{section_Topology} the electric field is referred to as $\vec{E}$. Furthermore, no magnetic field is applied; $\vec{B}=0$. Consequently,
\begin{align}
    \frac{\partial\vec{k}}{\partial t}=-\frac{e}{\hbar}\vec{\varepsilon}\quad\text{and}\quad\frac{\partial}{\partial t}=-\frac{e}{\hbar}\sum_i\varepsilon_i\frac{\partial}{\partial k_i}.
    \label{eq.tralalalala}
\end{align}
Inserting Equation \eqref{eq.tralalalala} into Equation \eqref{eq.gameboy} gives
\begin{align}
    \vec{v}_n&=\frac{1}{\hbar}\frac{\partial E_n}{\partial\vec{k}}+
    \frac{ie}{\hbar}\sum_i \varepsilon_i\left(
    \braket{\frac{\partial u_n}{\partial\vec{k}}|\frac{\partial u_n}{\partial k_i}}-
    \braket{\frac{\partial u_n}{\partial k_i}|\frac{\partial u_n}{\partial\vec{k}}}\right).
    \label{eq.Pizza}
\end{align}
Next we evaluate the $x$-component of the sum in Equation \eqref{eq.Pizza}:
\begin{align}
    &\sum_i \varepsilon_i
    \left(\braket{\frac{\partial u_n}{\partial k_x}|\frac{\partial u_n}{\partial k_i}}-\braket{\frac{\partial u_n}{\partial k_i}|\frac{\partial u_n}{\partial k_x}}\right)\\
    &=\varepsilon_x\cdot 0 + \varepsilon_y\underbrace{\left(\braket{\frac{\partial u_n}{\partial k_x}|\frac{\partial u_n}{\partial k_y}}-\braket{\frac{\partial u_n}{\partial k_y}|\frac{\partial u_n}{\partial k_x}}\right)}_{=\left[\bra{\frac{\partial u_n}{\partial\vec{k}}}\times\ket{\frac{\partial u_n}{\partial\vec{k}}}\right]_z}+
    \varepsilon_z\underbrace{\left(\braket{\frac{\partial u_n}{\partial\vec{k}_x}|\frac{\partial u_n}{\partial k_z}}-\braket{\frac{\partial u_n}{\partial k_z}|\frac{\partial u_n}{\partial k_x}}\right)}_{=-\left[\bra{\frac{\partial u_n}{\partial\vec{k}}}\times\ket{\frac{\partial u_n}{\partial\vec{k}}}\right]_y}\\
    &=\varepsilon_y\cdot\left[\bra{\frac{\partial u_n}{\partial\vec{k}}}\times\ket{\frac{\partial u_n}{\partial\vec{k}}}\right]_z-\varepsilon_z\cdot\left[\bra{\frac{\partial u_n}{\partial\vec{k}}}\times\ket{\frac{\partial u_n}{\partial\vec{k}}}\right]_y\\
    &= \left[\vec{\varepsilon}\times \left(\bra{\frac{\partial u_n}{\partial\vec{k}}}\times\ket{\frac{\partial u_n}{\partial\vec{k}}}\right)\right]_x
\end{align}
An analogous procedure for the $y$- and $z$-component gives
\begin{align}    
    \vec{v}_n &=\frac{1}{\hbar}\frac{\partial E_n}{\partial\vec{k}}+\frac{e}{\hbar}\vec{\varepsilon}\times\left(\vphantom{\frac{I}{I}}\right.\underbrace{i\bra{\frac{\partial u_n}{\partial\vec{k}}}\times\ket{\frac{\partial u_n}{\partial\vec{k}}}}_{\substack{\stackrel{\text{\eqref{eq.Ohio}}}{=}\vec{\Omega}_n}}\left.\vphantom{\frac{I}{I}}\right)\\
    &=\vec{v}_0+\frac{e}{\hbar}\vec{\varepsilon}\times\vec{\Omega}_n. \tag{\ref{eq.AnomalousVelocity}}
\end{align}
We derived Equation \eqref{eq.AnomalousVelocity}. In addition to the usual group velocity $\vec{v}_0=\partial\omega_n/\partial \vec{k}$ we obtained a component perpendicular to the electric field $\vec{\varepsilon}$ and to the Berry curvature $\Omega_n$, namely the anomalous velocity. Notice that $\Omega_n$ is here defined in $k$-space.
\end{appendices}

\newpage

\bibliography{sn-bibliography}

\end{document}